# Estimating Irving–Kirkwood pressure profiles from per-atom virial and density profiles: a closure for the contour-dependent term in planar liquid films

José L. Rivera[a)]
*Laboratorio de Modelamiento y Simulación Molecular, Facultad de Ciencias Físico-Matemáticas, Universidad Michoacana de San Nicolás de Hidalgo, Morelia 58030, Michoacán, México*
[a)]Electronic mail: jlrivera@umich.mx

## ABSTRACT

The local pressure tensor of an inhomogeneous fluid is not unique. Two tangential profiles are in common use in planar simulations: the Irving–Kirkwood (IK) profile, which requires a dedicated pass over every pair of atoms for every slab, and the per-atom virial (PAV) profile, which molecular-dynamics codes produce at negligible cost and coincides with the Harasima contour. The two differ by the derivative of a pair quantity, $P_T^{\mathrm{IK}} - P_T^{\mathrm{PAV}} = \mathrm{d}Q/\mathrm{d}z$. We show that, for Lennard-Jones free-standing films, $Q(z)$ can be predicted from the density profile and the bulk radial distribution function of the coexisting liquid: under the closure $\rho^{(2)} \simeq \rho(z_1)\rho(z_2)g(r_{12})$, with a pair-local interpolation of $g$ between its liquid and dilute limits, $Q$ is a double integral of $\rho(z)$ against a kernel built from $u'(r)g(r)$. The estimator $P_T^{\mathrm{IK,est}} = P_T^{\mathrm{PAV}} + \mathrm{d}Q/\mathrm{d}z$ is tested against IK profiles measured on the same trajectories for 20 films (thicknesses 7–25$\sigma$, three cutoff radii, two temperatures): the root-mean-square residual is 0.001–0.002 $\epsilon\sigma^{-3}$, against 0.010–0.021 for the uncorrected profile, and $Q$, the amount that the contour choice transfers between the two interfaces of a film with non-equivalent surfaces, is recovered to within 0.6% of the surface tension. Applied to the normal component, the same closure removes 83–92% of the spurious structure of the per-atom $zz$ profile, a test that needs no IK run. Its mixture form is validated on three two-component films with an adsorbate unequally distributed between their faces, where it resolves the tension of each face; a clean face recovers the tension of the pure film.

## I. INTRODUCTION

The pressure tensor of an inhomogeneous fluid is the central quantity of the mechanical route to interfacial thermodynamics. For a planar interface normal to $z$, the surface tension is the integral of the difference between the normal and tangential components, $\gamma = \int [P_N(z) - P_T(z)]\,\mathrm{d}z$ [1–3], and the local profiles themselves are used to locate the "mechanical" interface, to compute bending moments and the Tolman length [4,5], to characterize the interior of thin films and of confined fluids [6,7], and increasingly as a diagnostic of mechanical stability in free-standing liquid layers [8,9]. Yet the local pressure tensor is not unique. The microscopic momentum flux fixes only the divergence of the stress; any choice of the path ("contour") along which the force between two atoms is distributed in space yields a valid tensor, and the local profiles obtained with different contours differ [2,10–13]; the same ambiguity underlies the long discussion of the atomic-level stress in solids and molecular systems [14–19]. The reference choice is the Irving–Kirkwood (IK) contour [20,21], the straight line joining the two atoms; in a slab geometry the pair virial is spread uniformly over the slabs between $z_i$ and $z_j$.

The practical consequence of this choice is well known. What most molecular-dynamics codes compute at negligible cost is not a contour-based tensor but the per-atom virial, binned into slabs (the `stress/atom` compute in LAMMPS, for instance, is evaluated inside the force loop and is available on GPUs [22,23]); we call the result the per-atom virial (PAV) profile. Binning assigns the virial of each pair half to the slab of each atom; for the tangential components this is the weight of the Harasima contour [11,24], which is why PAV profiles are often called "Harasima profiles", but for the normal component it is not the weight of any contour (Sec. II.E), which is why we keep the two names apart and describe the measured profile by what is actually done to obtain it. The IK profile instead requires a separate pass over all pairs within the cutoff for every slab they cross; in LAMMPS this is done by `compute stress/cartesian` [25], which is not GPU-accelerated and in our runs builds its own neighbor list and evaluates the pair forces on the CPU, or by the method-of-planes compute `stress/mop/profile` [26–28]. In our own production runs (34 713 atoms, $r_c = 7.5\sigma$, pair forces on a GPU, profiles sampled every 200 steps) adding the IK compute raised the wall time of $10^6$ steps from 1.73 to 3.84 h, so that the IK pass took 55% of the run; each evaluation of the IK profile cost about 1.5 s, the equivalent of some 300 GPU force evaluations. The cost grows with the number of neighbors, i.e. with $r_c^3$, which is precisely the regime (large cutoffs, no tail corrections) in which free-standing LJ films are best studied [8,29,30], since the tail corrections to the pressure tensor and to $\gamma$ that make short cutoffs acceptable for a single thick interface [31–34] presuppose a bulk liquid on one side of it.

Two facts make the difference between the IK and PAV tangential profiles tractable. First, because both distribute the same pair virial along $z$ with normalized weights, they differ by a total derivative, $P_T^{\mathrm{IK}}(z) - P_T^{\mathrm{PAV}}(z) = \mathrm{d}Q/\mathrm{d}z$, the general property of contour-dependent terms established by Schofield and Henderson [10], so that integrated quantities such as $\gamma$ are invariant while local values are not. Second, $Q(z)$ has an explicit pair form: each pair contributes a linear ramp between $z_i$ and $z_j$. $Q$ is thus a two-body quantity of exactly the same kind as the pressure itself, and it can be approximated with the same closures used for the pressure in the theory of inhomogeneous fluids. The simplest of these, $\rho^{(2)}(\mathbf{r}_1, \mathbf{r}_2) \simeq \rho(z_1)\rho(z_2)g(r_{12})$, is known to give a poor absolute $P_T(z)$ at a liquid–vapor interface [2,3,11]; what we show here is that it gives a very good $\mathrm{d}Q/\mathrm{d}z$, because the contour difference is dominated by the geometry of the pairs that straddle a given plane, which the closure captures, and not by the subtle correlations that fix the absolute value of the pressure, which it does not.

In this paper we (i) write $Q$ for the IK–PAV pair in closed form and reduce it, under the closure, to a double integral of the density profile against a kernel built from $u'(r)g(r)$; (ii) introduce a pair-local interpolation of $g(r)$ between its liquid and dilute limits that is needed at the interface; (iii) test the resulting estimator $P_T^{\mathrm{IK,est}} = P_T^{\mathrm{PAV}} + \mathrm{d}Q/\mathrm{d}z$, with $P_T^{\mathrm{PAV}}$ the measured PAV tangential profile, against IK profiles measured on the same trajectories for 20 LJ films covering thicknesses from 7 to $25\sigma$, three cutoff radii and two temperatures; and (iv) state explicitly which components of the PAV profile correspond to a legitimate contour and which do not. The practical outcome is that the IK tangential profile can be obtained from a cheap PAV run plus the density profile and one short bulk simulation of the coexisting liquid, with a residual in the liquid interior that is within a factor of two of the statistical noise at $T = 0.72$ and at the noise level at $T = 0.90$. We stress that the total surface tension does not need any of this: it is contour invariant and the PAV profile gives it correctly, as we verify below for every film. The local profile is needed when the question is local. The case that motivates us is that of films with two non-equivalent interfaces, for instance a liquid film with a second component adsorbed at only one of its surfaces, where the quantity of interest is the tension of each interface separately; these are obtained by splitting $\int (P_N - P_T)\,\mathrm{d}z$ at a dividing plane inside the film, and the split is contour dependent unless the dividing plane lies in a bulk-like region, which a thin film does not have. The same applies to the surface of tension, to the stress-based stability criteria of Refs. [8,9], and to any comparison of the two sides of a film.

## II. THEORY

### A. The IK and PAV weights and the gauge term

Consider a planar system with area $A$ in the $xy$ plane, periodic in $x$ and $y$, with atoms interacting through a pair potential $u(r)$. Any slab-resolved tangential pressure built from pair virials, averaged over $x$ and $y$ and over the two in-plane directions, can be written as

$$P_T^C(z) = k_B T \rho(z) - \frac{1}{A}\left\langle \sum_{i<j} v_{ij}\ w_C\big(z; z_i, z_j\big) \right\rangle, \tag{1}$$

with $v_{ij} = \left[\left(x_{ij}^2 + y_{ij}^2\right)/2r_{ij}\right] u'\left(r_{ij}\right)$, where $\rho(z)$ is the number density, $r_{ij} = \left|\mathbf{r}_i - \mathbf{r}_j\right|$, $v_{ij}$ is the tangential virial of the pair, and $w_C$ is a normalized weight, $\int w_C \, \mathrm{d}z = 1$, that specifies how the virial of the pair is distributed along $z$; the label $C$ stands for the prescription used. For the IK contour the weight is uniform between the two atoms,

$$w_{\mathrm{IK}}\left(z; z_i, z_j\right) = \frac{1}{\left|z_{ij}\right|} \Theta\left(\frac{z - z_i}{z_{ij}}\right) \Theta\left(\frac{z_j - z}{z_{ij}}\right), \tag{2}$$

with $z_{ij} = z_j - z_i$ and $\Theta$ the Heaviside function, whereas binning the per-atom virial splits it between the planes of the two atoms,

$$w_{\mathrm{PAV}}\left(z; z_i, z_j\right) = 1/2\, \delta(z - z_i) + 1/2\, \delta\left(z - z_j\right), \tag{3}$$

since the per-atom virial $\mathbf{S}_i = -1/2 \sum_j \mathbf{r}_{ij} \otimes \mathbf{f}_{ij}$ carries half of every pair term. Equation (3) is also the tangential weight of the Harasima contour [11,24], so $P_T^{\mathrm{PAV}} = P_T^{\mathrm{H}}$ and everything below applies to the Harasima profile as well; the normal component is discussed in Sec. II.E. In mechanical equilibrium the IK normal pressure $P_N(z)$ is constant and equal to the vapor pressure. Figure 1 sketches the two weights.

Because both weights are normalized, their difference integrates to zero and can be written as the derivative of a bounded function,

$$w_{\mathrm{IK}} - w_{\mathrm{PAV}} = \frac{\mathrm{d}q}{\mathrm{d}z}, \quad q\left(z; z_i, z_j\right) = \int_{-\infty}^{z} \left(w_{\mathrm{IK}} - w_{\mathrm{PAV}}\right) \mathrm{d}z'. \tag{4}$$

From Eqs. (2) and (3), $q$ is a linear ramp that vanishes outside $\left[z_i, z_j\right]$ and runs from $-1/2$ just above the lower atom to $+1/2$ just below the upper one [Fig. 1(c)]. Inserting (4) in (1) gives the difference between the two profiles as a total derivative, the form that Schofield and Henderson [10] showed to hold between any two admissible contours, here in explicit pair form,

$$P_T^{\mathrm{IK}}(z) - P_T^{\mathrm{PAV}}(z) = \frac{\mathrm{d}Q}{\mathrm{d}z}, \quad Q(z) = -\frac{1}{A}\left\langle \sum_{i<j} v_{ij}\, q_{ij}(z) \right\rangle, \tag{5}$$

with $q_{ij}(z) \equiv q\left(z; z_i, z_j\right)$. Since $q$ vanishes far from the interface in a homogeneous region and is odd about the midpoint of each pair, $Q$ vanishes in bulk phases and $\int (\mathrm{d}Q/\mathrm{d}z)\, \mathrm{d}z = 0$ across the whole system: the surface tension of a film with two interfaces, $\gamma = 1/2 \int (P_N - P_T)\, \mathrm{d}z$ over the whole cell, and every other integral of $P_T$ are the same for the two profiles, as they must be, while the local values are not. The kinetic term drops out of the difference. Partial integrals are not invariant: the tension assigned to the interface below a dividing plane $z_s$, $\gamma_1 = \int_{-\infty}^{z_s} (P_N - P_T)\, \mathrm{d}z$, differs between the two profiles by $\gamma_1^{\mathrm{IK}} - \gamma_1^{\mathrm{PAV}} = -Q(z_s)$, and the complementary $\gamma_2$ by $+Q(z_s)$. $Q$ vanishes wherever the fluid is homogeneous over a distance $r_c$ on both sides, so the split is invariant for a thick film with the plane in its bulk-like interior, and also for a symmetric film split at its midplane, where $Q$ vanishes by symmetry; for a thin film with non-equivalent interfaces neither condition holds, and the choice of profile transfers an amount $Q(z_s)$ of tension from one interface to the other. Equation (5) is exact for any pair potential; it is the quantity we set out to estimate, and its value at the dividing plane is precisely what a per-interface analysis needs.

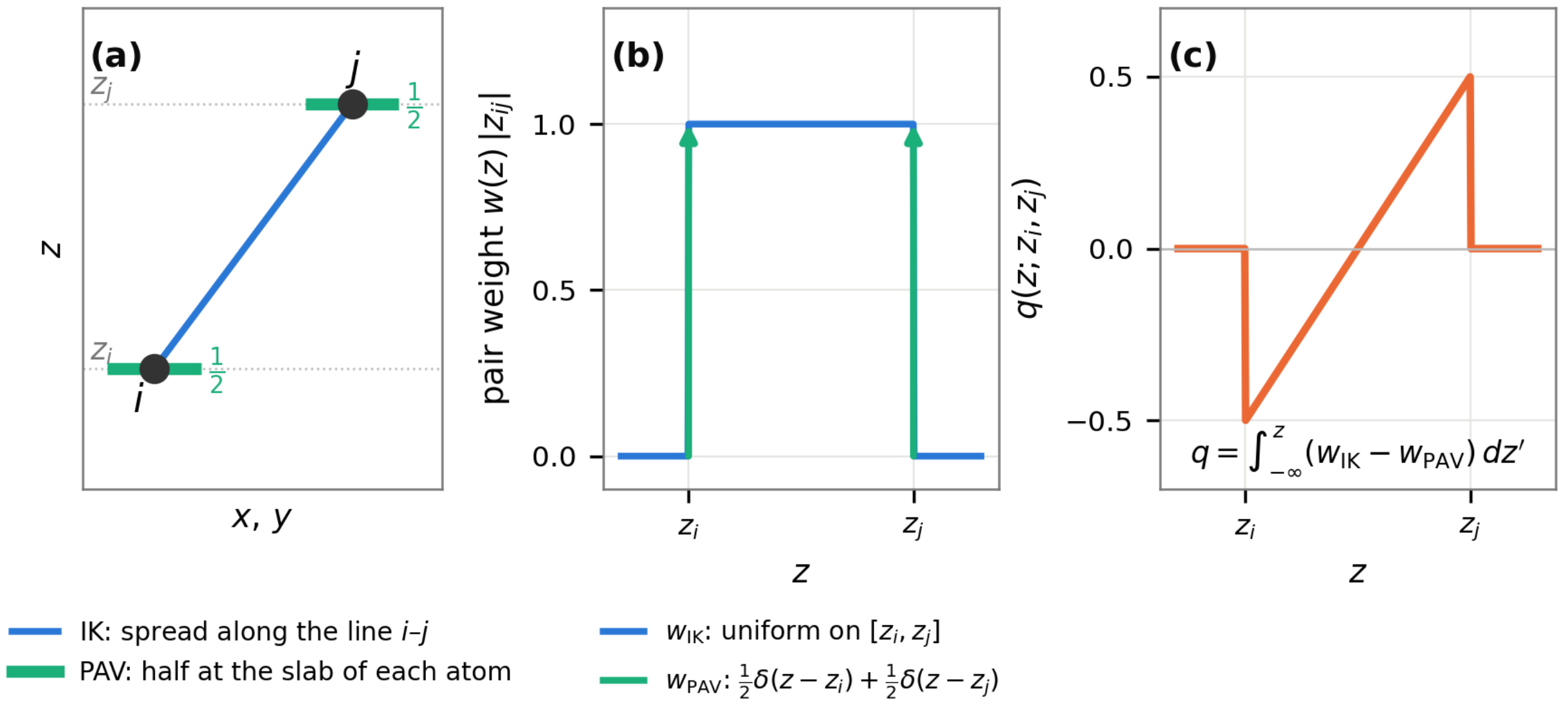

***Figure 1.*** *(a) A pair of atoms i, j and the two prescriptions for locating its tangential virial along z: the IK contour spreads it along the straight line between the atoms, per-atom virial binning (PAV) places half of it in the slab of each atom. (b) The corresponding weights, Eqs. (2) and (3). (c) Their cumulative difference* $q(z; z_i, z_j)$*, Eq. (4): a ramp from* $-1/2$ *to* $+1/2$ *confined to* $[z_i, z_j]$*. The gauge term* $Q(z)$ *of Eq. (5) is the virial-weighted sum of these ramps over all pairs.*

## B. Closure

Writing the pair sum in (5) as an integral over the pair density $\rho^{(2)}(\mathbf{r}_1, \mathbf{r}_2)$ and adopting the closure

$$\rho^{(2)}(\mathbf{r}_1, \mathbf{r}_2) \simeq \rho(z_1)\,\rho(z_2)\,g(r_{12}), \tag{6}$$

with $g(r)$ the radial distribution function of the homogeneous liquid at coexistence, the in-plane integration can be carried out once and for all. With $r^2 = s^2 + \zeta^2$, $\zeta = z_2 - z_1$, and $2\pi s\,\mathrm{d}s = 2\pi r\,\mathrm{d}r$, the factor $(x^2 + y^2)/2r = s^2/2r$ integrates to

$$K(\zeta) = \pi \int_{|\zeta|}^{r_c} (r^2 - \zeta^2)\, u'(r)\, g(r)\, \mathrm{d}r, \tag{7}$$

which is an even function of $\zeta$ with support $|\zeta| < r_c$ for a truncated potential. In terms of $K$, the configurational parts of the two tangential profiles, and $Q$ itself, become functionals of the density profile alone:

$$P_T^{\mathrm{PAV},V}(z) = -1/2 \int \mathrm{d}\zeta\, K(\zeta)\, \rho(z)\rho(z+\zeta), \tag{8}$$

$$P_T^{\mathrm{IK},V}(z) = -1/2 \int \mathrm{d}\zeta\, K(\zeta) \int_0^1 \mathrm{d}\,\alpha \times \rho(z - \alpha\zeta)\, \rho(z + (1-\alpha)\zeta), \tag{9}$$

$$Q(z) = -1/2 \int \mathrm{d}\zeta\, |\zeta|\, K(\zeta) \int_0^1 \mathrm{d}\,\alpha\, (\alpha - 1/2) \times \rho(z - \alpha\zeta)\, \rho(z + (1-\alpha)\zeta). \tag{10}$$

Here $\alpha$ is the fractional position of the plane $z$ along the pair, and the factor $|\zeta|$ in (10) is the Jacobian of the change of variable from $z_1$ to $\alpha$. In a homogeneous fluid of density $\rho$, Eqs. (8) and (9) both reduce to $-1/2\,\rho^2 \int K\,\mathrm{d}\zeta = -2\pi/3\,\rho^2 \int_0^{r_c} r^3\,u'(r)g(r)\,\mathrm{d}r$, the configurational virial pressure, and (10) vanishes by symmetry. We note that the gradient expansion of (10), $Q \simeq -\pi/45\,\rho\rho' \int r^5 u' g\,\mathrm{d}r$, is not useful here: it requires $\rho$ to vary slowly over the range of $K$, whereas the interfacial width of the LJ liquid–vapor interface ($\approx 2\sigma$ at $T^* = 0.72$) is much smaller than $r_c$. All results below use the full non-local form.

## C. The estimator

The closure (6) is a crude approximation for the absolute profiles: at a liquid–vapor interface it misses the anisotropic correlations that produce the deep tangential minimum, and in our systems the surface tension computed from Eqs. (8) or (9) alone is too low by a factor of 1.4–3. We therefore do not use the closure for the pressure. We use it only for the difference between the two profiles, and define the estimator

$$P_T^{\mathrm{IK,est}}(z) = P_T^{\mathrm{PAV}}(z) + \frac{\mathrm{d}Q}{\mathrm{d}z}, \tag{11}$$

where $P_T^{\mathrm{PAV}}$ is the tangential PAV profile measured in the simulation and $Q$ is the closure expression (10). The rationale is that $\mathrm{d}Q/\mathrm{d}z$ is controlled by how many pairs straddle the plane $z$ and by how asymmetrically they do so, which is a statement about the density on both sides of $z$ within a distance $r_c$, whereas the absolute pressure is controlled by the detailed pair structure at $r \approx \sigma$. The former is what (6) retains; the latter is what it loses.

*Pair-local* $g(r)$. With a single liquid $g(r)$ the closure overestimates the structure of pairs that have one or both members in the vapor. We therefore let the kernel depend on the local environment of the pair through a linear interpolation between a liquid kernel $K_\ell$, built with the bulk-liquid $g(r)$ at coexistence, and a dilute kernel $K_g$, built with the low-density limit $g(r) = \exp[-u(r)/k_B T]$:

$$K_{12} = w_{12} K_\ell + (1 - w_{12}) K_g, \quad w_{12} = \min[1, \bar{\rho}_{12}/\rho_\ell], \tag{12}$$

with $\bar{\rho}_{12} = 1/2\,[\rho(z_1) + \rho(z_2)]$ and $\rho_\ell$ the coexistence liquid density. Equation (12) is applied inside the $\zeta$ and $\alpha$ integrals of (10), so that a pair is weighted by the mean density at the two positions it actually occupies. This refinement has no effect in the liquid interior and reduces the residual at the interface by roughly a factor of two (Sec. IV).

The inputs of the estimator are therefore the measured PAV tangential profile, the density profile of the same run (or, equivalently, the per-bin atom counts that the binning compute already produces), the pair potential, and $g(r)$ of the bulk liquid at the same temperature and cutoff. Nothing else is needed; in particular no additional per-pair pass over the film trajectory.

## D. Mixtures

Everything above is written pair by pair and carries over to a mixture without change of structure. The weights (2) and (3) and the ramp (4) do not depend on the species of $i$ and $j$, so Eq. (5) holds with the sum running over all pairs, and the measured PAV and IK profiles of a mixture are the ones the codes produce with no modification. The closure is applied to each pair of species,

$$\rho_{\alpha\beta}^{(2)}(\mathbf{r}_1, \mathbf{r}_2) \simeq \rho_\alpha(z_1)\,\rho_\beta(z_2)\,g_{\alpha\beta}(r_{12}), \tag{13}$$

with $\rho_\alpha(z)$ the partial density profiles, and the gauge term becomes a sum over ordered species pairs,

$$Q(z) = \sum_{\alpha\beta} Q_{\alpha\beta}\,(z), \tag{14}$$

where $Q_{\alpha\beta}$ is Eq. (10) with $\rho(z-\alpha\zeta) \to \rho_\alpha(z-\alpha\zeta)$, $\rho(z+(1-\alpha)\zeta) \to \rho_\beta(z+(1-\alpha)\zeta)$, and the kernel $K_{\alpha\beta}$ of Eq. (7) built from $u'_{\alpha\beta}(r) g_{\alpha\beta}(r)$; the factor $1/2$ of (10) together with the sum over ordered pairs counts each physical pair once, as in the pure case. The same replacement in the normal kernel of Sec. II.F gives $Q_N$. The estimator (11) is unchanged: the measured PAV profile of the mixture plus $\mathrm{d}Q/\mathrm{d}z$ from (14).

The only new input is $g_{\alpha\beta}(r)$. When the mixture has a bulk liquid phase of the relevant composition, the three (for a binary) radial distribution functions can be taken from one bulk run, exactly as in the pure case. In the case that motivates this work, a dilute component that is surface active and nearly absent from the interior, there is no bulk reference for the cross and minority pairs, and we use the conformal approximation

$$g_{\alpha\beta}(r) \simeq g(r/\sigma_{\alpha\beta}), \tag{15}$$

with $g$ the liquid $g(r)$ of the majority species, mixed locally with the dilute limit $\exp[-u_{\alpha\beta}(r)/k_B T]$ of each pair as in Eq. (12), with $w_{12}$ evaluated with the total density $\rho = \sum_\alpha \rho_\alpha$. Two remarks justify the choice. First, as discussed below (Sec. V), $Q$ is dominated by pairs with $|\zeta|$ of the order of $r_c$, for which $g \simeq 1$ and the kernel is insensitive to the details of $g$; the approximation (15) therefore affects $Q_{\alpha\beta}$ only through the short pairs, which for a dilute adsorbate are a small fraction of the whole. Second, the scaling of the argument is not optional: the unscaled $g$ of the majority species would place the cross and minority pairs inside their repulsive cores, where $u'_{\alpha\beta}$ is large, and in our tests it produced errors in $Q$ far larger than those of the pure films; with (15) the core of each pair is at $r \simeq \sigma_{\alpha\beta}$ and the accuracy of the pure case is recovered (Sec. IV.J). The practical inputs of the mixture estimator are thus the PAV profile, the per-species density profiles (the per-group chunk averages of the same run), the pair parameters, and one bulk $g(r)$. The reference implementation accepts any number of species and pair parameters.

For a film with non-equivalent interfaces the quantity of interest is the tension of each face, $\gamma_k = \int_k (P_N - P_T)\,\mathrm{d}z$ over the half of the cell that contains face $k$, split at a dividing plane $z_s$ inside the film. By Eq. (5), $\gamma_1^{\mathrm{IK}} - \gamma_1^{\mathrm{PAV}} = -Q(z_s) = -(\gamma_2^{\mathrm{IK}} - \gamma_2^{\mathrm{PAV}})$: the two contours redistribute the amount $Q(z_s)$ between the faces, the total being invariant, and the estimator supplies both the IK partition and the transferred amount from a PAV run.

## E. The normal component of the PAV profile

A point that is frequently blurred in practice deserves to be stated explicitly, and it is the reason we do not call the measured profiles "Harasima profiles". Binning the per-atom virial assigns the virial of each pair half to the slab of $i$ and half to the slab of $j$ for all tensor components, so the weight (3) applies to $P_{zz}^{\mathrm{PAV}}$ as well. No admissible contour does this for the normal component [35]: a contour must distribute $z_{ij} f_{z,ij}$ between the two atoms so that $P_N$ satisfies the mechanical-equilibrium condition, and the Harasima contour, whose tangential weight is (3), spreads it uniformly between $z_i$ and $z_j$ exactly as IK does [11,36]. $P_N^{\mathrm{PAV}}(z)$ is therefore not the normal component of the Harasima contour or of any other; it exhibits a structured profile across the interface that does not satisfy the mechanical-equilibrium condition $P_N = \mathrm{const}$ [36,37]. The IK normal component, or equivalently the method of planes [26], must be used for $P_N$. Its integral over $z$, however, is correct: $\int P_N^{\mathrm{PAV}}\,\mathrm{d}z = \int P_N^{\mathrm{IK}}\,\mathrm{d}z$, since both equal the total $zz$ virial per unit area, so a surface tension computed entirely from PAV profiles is correct as well; only the local shape of $P_N^{\mathrm{PAV}}$ is meaningless. This is immaterial for the tangential estimator, but it matters for any criterion built on the anisotropy $P_N - P_T$: the PAV anisotropy is contaminated at the interface by a spurious $P_N^{\mathrm{PAV}}$, while its integral (the surface tension) is correct. All anisotropies reported below use the IK (constant) normal component, which in our single-component films equals the vapor pressure to within noise throughout the system.

## F. The normal component: a test of the closure

The construction of Sec. II.A is not specific to the tangential component. With $v_{ij}^N = (z_{ij}^2/r_{ij})\,u'(r_{ij})$ in place of $v_{ij}$, Eq. (1) gives the normal pressure for any weight, the PAV weight (3) applies to $zz$ as well, and the same ramp $q_{ij}$ yields

$$P_N^{\mathrm{IK}} - P_N^{\mathrm{PAV}} = \frac{\mathrm{d}Q_N}{\mathrm{d}z}, \quad Q_N = -\frac{1}{A}\left\langle \sum_{i<j} v_{ij}^N \, q_{ij} \right\rangle. \tag{16}$$

Under the closure (6), $Q_N$ is given by Eq. (10) with the kernel

$$K_N(\zeta) = 2\pi\zeta^2 \int_{|\zeta|}^{r_c} u'(r)\, g(r)\, \mathrm{d}r, \tag{17}$$

which has the same integral as $K$ (both give the bulk virial) but a very different shape: $K$ is largest at $\zeta = 0$ and is fed by in-plane pairs, including the short ones for which $g(r)$ matters, whereas $K_N$ vanishes at $\zeta = 0$ and is dominated by pairs with $|\zeta|$ of the order of $r_c$, for which $g \simeq 1$. Two consequences follow. First, Eq. (16) identifies the structured $P_N^{\mathrm{PAV}}(z)$ of Sec. II.E as $-\mathrm{d}Q_N/\mathrm{d}z$ plus a constant, i.e. as a pure gauge artifact of the per-atom assignment. Second, since $P_N^{\mathrm{IK}}$ is flat and equal to $P_{\mathrm{vap}}$ in mechanical equilibrium, the corrected profile $P_N^{\mathrm{PAV}} + \mathrm{d}Q_N/\mathrm{d}z$ must be flat: this is a test of the closure that requires no IK run at all, only the $zz$ column of the per-atom stress, and its residual non-flatness measures the closure error directly, for every system, with a known answer. The weighting of $K_N$ towards long pairs might be expected to favour the closure; the test (Sec. IV.H) shows a capture comparable with, not better than, the tangential one, because the short pairs that straddle the interface are the ones for which the closure fails in both cases.

# III. METHODS

## A. Systems

We simulated free-standing films of a single-site LJ fluid, $u(r) = 4\epsilon[(\sigma/r)^{12} - (\sigma/r)^6]$ truncated at $r_c$ without shift or long-range corrections, in a cell of $49.6\sigma \times 49.6\sigma \times 48\sigma$ periodic in all three directions, with the film normal to $z$ and vapor on both sides. Reduced units are used throughout ($T^* = k_B T/\epsilon$, $\rho^* = \rho\sigma^3$, pressures in $\epsilon\sigma^{-3}$); the asterisk is dropped below. The cell and protocol follow our earlier work on these films [8,29], which in turn follows the classical setup of planar liquid–vapor simulations of the LJ fluid [32,38–40], with the lateral size $b_t = 49.6\sigma$ kept fixed so that the finite-size effects on $\gamma$ and $\rho_\ell$ that depend on the lateral size [41,42] are the same for all systems; the dependence on the film thickness [6,29,43] is part of what the thickness series probes. Three series were run (Table 1): a thickness series at $T = 0.72$, $r_c = 7.5\sigma$ (seven films, $\ell_t = 7.0$ to $25.0\sigma$); a cutoff series at $T = 0.72$ with $r_c = 3.5\sigma$ (three films) and $r_c = 5.0\sigma$ (five films); and a temperature series at $T = 0.90$, $r_c = 7.5\sigma$ (five films, $\ell_t = 9.3$ to $17.6\sigma$). Thinner films at each cutoff ($\ell_t \lesssim 5\sigma$ at $r_c = 7.5$, $6\sigma$ at $r_c = 3.5$) ruptured during equilibration and are not included. The thickness $\ell_t$ reported is the distance between the two half-density points of $\rho(z)$.

*Metastability of the films in a periodic cell.* A remark on the cell geometry is in order, because it is a recurrent objection to simulations of this kind. In a periodic cell of dimensions $b_t \times b_t \times L_z$ a planar film normal to $z$ is the state of lowest interfacial free energy only if its area, $2b_t^2$, is the smallest that the liquid volume can adopt among the competing morphologies of the finite system: a slab normal to $x$ or $y$ (area $2b_tL_z$), a cylinder ($2\pi R b_t$) and a droplet ($4\pi R^2$) [44–46]. With $b_t \simeq L_z$ as here, a slab normal to $x$ has 3% less area than one normal to $z$ for every thickness, and for $\ell_t \lesssim 16\sigma$ a cylinder, and for $\ell_t \lesssim 13\sigma$ a droplet, have less area still (for $\ell_t = 7\sigma$, 3280 and $3230\sigma^2$ against $4920\sigma^2$). Strictly, therefore, most of the films of Table 1 are metastable, as are the free-standing films of experiment. They are nevertheless well-defined stationary states on any simulation time scale, because the transformation to a lower-area morphology requires a hole to be nucleated through the film, with a free-energy barrier of the order of that of the classical critical hole of radius $\ell_t/2$, $\Delta G^* \simeq \pi\gamma\ell_t^2/2$ [47], i.e. $\approx 110\, k_B T$ for $\ell_t = 7\sigma$ and growing as $\ell_t^2$; the corresponding rate is unobservable. Consistently, none of the 23 films changed its morphology during its run ($10^5\tau$ for the single-component films, $2 \times 10^4\tau$ for the two-component ones); the thinner films that ruptured during equilibration (see above) are outside the scope of this work, and the mechanism of their rupture is not analysed here. In a previous work [29] we also verified for a $16.8\sigma$ film with $b_t = 150.4\sigma$ that lengthening the cell from $L_z = 48\sigma$ to $288\sigma$ (aspect ratios from 3.1 to 0.5) leaves the density and pressure profiles, $\gamma$ and its variance unchanged. The thermodynamic criterion of Singh *et al.* [48] for the spontaneous formation of a planar interface from a homogeneous system (the interfacial area must be smaller than a critical area set by the van der Waals loop) addresses a different question, the competition with the homogeneous state, and is satisfied by a wide margin here: at the mean density of any of our cells ($\bar{\rho} = N/V = 0.12$–$0.43$ at $T = 0.72$, $0.16$–$0.28$ at $T = 0.90$) the homogeneous fluid is not metastable but unstable, as we verified directly by letting homogeneous systems at these densities, temperatures and cutoffs evolve in the canonical ensemble: they decompose without incubation, the coarse-grained density contrast doubling before the first sampled configuration, at $12.5\tau$, and reaching liquid-like local densities within $50\tau$ (supplementary material, Sec. S8). For the purpose of this work, finally, the question is immaterial: Eq. (5) is an identity for every configuration, both profiles are accumulated on the same configurations, and the closure is evaluated with the density profile of the film actually simulated, so the test of the estimator is valid for any planar film, stable, metastable or even slowly relaxing (Sec. IV.J).

To test the method on films with non-equivalent interfaces we also simulated three two-component systems in the same cell at $T = 0.72$, $r_c = 7.5\sigma$: a film of 20 886 atoms of the LJ liquid (species A, $\ell_t = 10.1$–$10.2\sigma$) and 622 atoms of a larger, weakly self-interacting adsorbate (species B; $\epsilon_{BB} = 0.5\epsilon$, equal masses, all pairs truncated at $r_c$), initially placed on the upper face only at about half a monolayer. Film I has $\sigma_{BB} = 1.5\sigma$, $\sigma_{AB} = 1.25\sigma$, $\epsilon_{AB} = 0.7\epsilon$; film II has $\sigma_{BB} = 2.0\sigma$, $\sigma_{AB} = 1.5\sigma$, $\epsilon_{AB} = 0.7\epsilon$; film III has the diameters of film II and $\epsilon_{AB} = 0.5\epsilon$. The B particles are surface active because their binding to the surface of A exceeds what they gain by dissolving or by evaporating. In films I and II they are slightly soluble in A ($x_B \simeq 5 \times 10^{-4}$ and $9 \times 10^{-5}$ in the interior), and over the $2 \times 10^4\tau$ of the runs part of them diffused through the film to the lower face: 30% in film I and 11% in film II, most of the latter during the short melt at $T = 0.95$. In film III the weaker A–B attraction makes B insoluble ($x_B \simeq 4 \times 10^{-6}$; a single B particle reached the lower face) but more volatile, with 24% of B in the vapor above the decorated face; since the wall on B confines that vapor to the upper gap, the normal pressure is $0.0033\,\epsilon\sigma^{-3}$ higher in the upper vapor than in the lower one, the difference being the partial pressure of B that the wall sustains. The adsorbed coverages (B within $5\sigma$ of the half-density point of each face) are $\Gamma_1 = 0.071$, $\Gamma_2 = 0.159\sigma^{-2}$ for film I, 0.029, $0.205\sigma^{-2}$ for film II and 0, $0.193\sigma^{-2}$ for film III, with 8–9% of B in the vapor of films I and II. Films I and II show a slow residual drift during production (175/447 and 199/423 B particles in the two halves of the cell for the two halves of the run of film I, 66/556 and 75/547 in film II), so that they are snapshots of slowly relaxing systems rather than equilibrium states; film III is stationary. The drift does not affect the test of the estimator, which compares two stress profiles accumulated on the same configurations, and the per-interface tensions obtained from it are the tensions of the time-averaged coverages. The equilibration of these systems was $2 \times 10^6$ steps after a melt of $10^5$ steps at $T = 0.95$, the production $2 \times 10^6$ steps, and a harmonic wall acting on B alone inside the vapor gap prevented B from crossing the periodic boundary; the LAMMPS input is supplied as supplementary material.

## B. Simulation protocol

Simulations were run with LAMMPS [23] in the canonical ensemble with a Nosé–Hoover thermostat, time step $\delta t = 0.005$, and the centre of mass of the film re-centred in $z$ at every step (`fix recenter`). Films were equilibrated for $2 \times 10^7$ steps (all three series), followed by production runs of $2 \times 10^6$ steps during which the two profiles were accumulated on the same trajectory, sampled on the same time steps, in slabs of width $0.5\sigma$: the PAV profile, from `compute stress/atom` binned with `fix ave/chunk`, and the IK tensor with `compute stress/cartesian` [25]. The normal pressure is taken from the IK tensor; the method of planes [26] was not used, because of its cost and because the IK normal pressure is already flat and equal to the vapor pressure to within noise (Sec. IV.H). The density profile was taken from the atom counts of the chunk average. Profiles were accumulated in blocks of $10^4$ steps; the standard error per bin quoted below is the standard error of the block means of the difference $P_T^{\mathrm{IK}} - P_T^{\mathrm{PAV}}$, which is the quantity against which the estimator is compared.

### C. Bulk $g(r)$ and evaluation of the closure

For each $(T, r_c)$ pair, $g(r)$ of the bulk liquid was computed in a separate NVT simulation of 4000 atoms at the coexistence density $\rho_\ell$ measured in the interior of the thickest film of the series ($\rho_\ell \simeq 0.81$, 0.83, 0.83 for $r_c =$ 3.5, 5.0, 7.5 at $T = 0.72$; 0.75 at $T = 0.90$), with the same truncated potential, $5 \times 10^4$ steps of equilibration and $10^5$ steps of sampling with `compute rdf` (750 bins up to $r_c$). These runs cost a negligible fraction of a film run. The kernel (7) was tabulated on $\zeta \in [-r_c, r_c]$ with $\Delta r = 0.002\sigma$; the integrals (10) were evaluated on a fine grid ($\Delta z = 0.1\sigma$) with the density profile interpolated linearly and periodically, a 64-point midpoint rule in $\alpha$, and then resampled onto the $0.5\sigma$ bins of the simulation. $\mathrm{d}Q/\mathrm{d}z$ was obtained by centred differences on the fine grid. The whole evaluation takes a few seconds in Python for a 96-bin profile. For the two-component films the closure was evaluated as in Sec. II.D, Eqs. (14) and (15), over the four ordered species pairs, with the partial density profiles from per-species chunk averages (`fix ave/chunk` on each group, `density/number`) and the $g(r)$ of the pure liquid at $r_c = 7.5\sigma$ for all pairs. A reference implementation `gauge.py` is provided as supplementary material.

### D. Observables and statistics

For each film we report (i) the full profiles; (ii) the residual $r(z) = P_T^{\mathrm{IK,meas}} - P_T^{\mathrm{IK,est}}$ and its root-mean-square (rms) over all bins, compared with the rms of the uncorrected difference $P_T^{\mathrm{IK,meas}} - P_T^{\mathrm{PAV}}$; (iii) the same quantities restricted to the liquid interior ($\rho > 0.9\rho_{\max}$), to the interfacial bins (the remaining bins within $2\sigma$ of the half-density points) and to the vapor; (iv) in the liquid, $\chi^2$ per bin, i.e. the ratio of the mean-squared residual to the mean-squared standard error of the measured difference; and (v) the central-plane anisotropy

$$\Delta P_T = P_{\mathrm{vap}} - P_T(z_c), \tag{18}$$

with $P_T(z_c)$ averaged over $|z - z_c| \leq 1\sigma$ and $P_{\mathrm{vap}}$ the IK pressure in the vapor more than $r_c + 2\sigma$ away from either interface; (vi) the surface tension $\gamma = 1/2 \sum_k \left(P_N^{\mathrm{IK}} - P_T\right)_k \Delta z$ summed over all bins of the cell, evaluated with the IK, PAV and estimated tangential profiles, with its statistical error from the block series of the difference; and (vii) for the two-component films, the per-interface tensions $\gamma_1, \gamma_2$ of Sec. II.D, with the dividing plane at the bin edge nearest the film centre (the midpoint of the half-density points of $\rho_A$), for the three tangential profiles, together with the measured $Q(z_s)$, i.e. the integral of $P_T^{\mathrm{IK}} - P_T^{\mathrm{PAV}}$ up to $z_s$, and its closure estimate. $\Delta P_T$ vanishes for a film whose interior is mechanically bulk-like and is the natural observable of the mechanical-stability criterion of Refs. [8,9]; because it is a local value, it is contour dependent, and the estimator is required precisely to obtain its IK value from a PAV run.

## IV. RESULTS

### A. Profiles

Figure 2 shows the complete picture for the thinnest and the thickest films of the $T = 0.72$, $r_c = 7.5\sigma$ series. The measured PAV and IK tangential profiles [panels (a) and (c)] share the same deep minima at the interfaces but differ systematically: the PAV minima are broader and shallower, the IK ones narrower and deeper, and in the $7\sigma$ film the IK profile lies below the PAV one at the centre by 0.028. The difference [panels (b) and (d)] is an antisymmetric doublet at each interface, positive on the vapor side and negative on the liquid side, with a magnitude of $\pm 0.05$–$0.07$ that is comparable with the vapor pressure times 30 and with the central anisotropy of the thinnest films; it is not a small correction. The closure prediction $\mathrm{d}Q/\mathrm{d}z$ reproduces this doublet in position, width and amplitude, and the estimated IK profile (solid line in (a) and (c)) is indistinguishable from the measured one on the scale of the full profile.

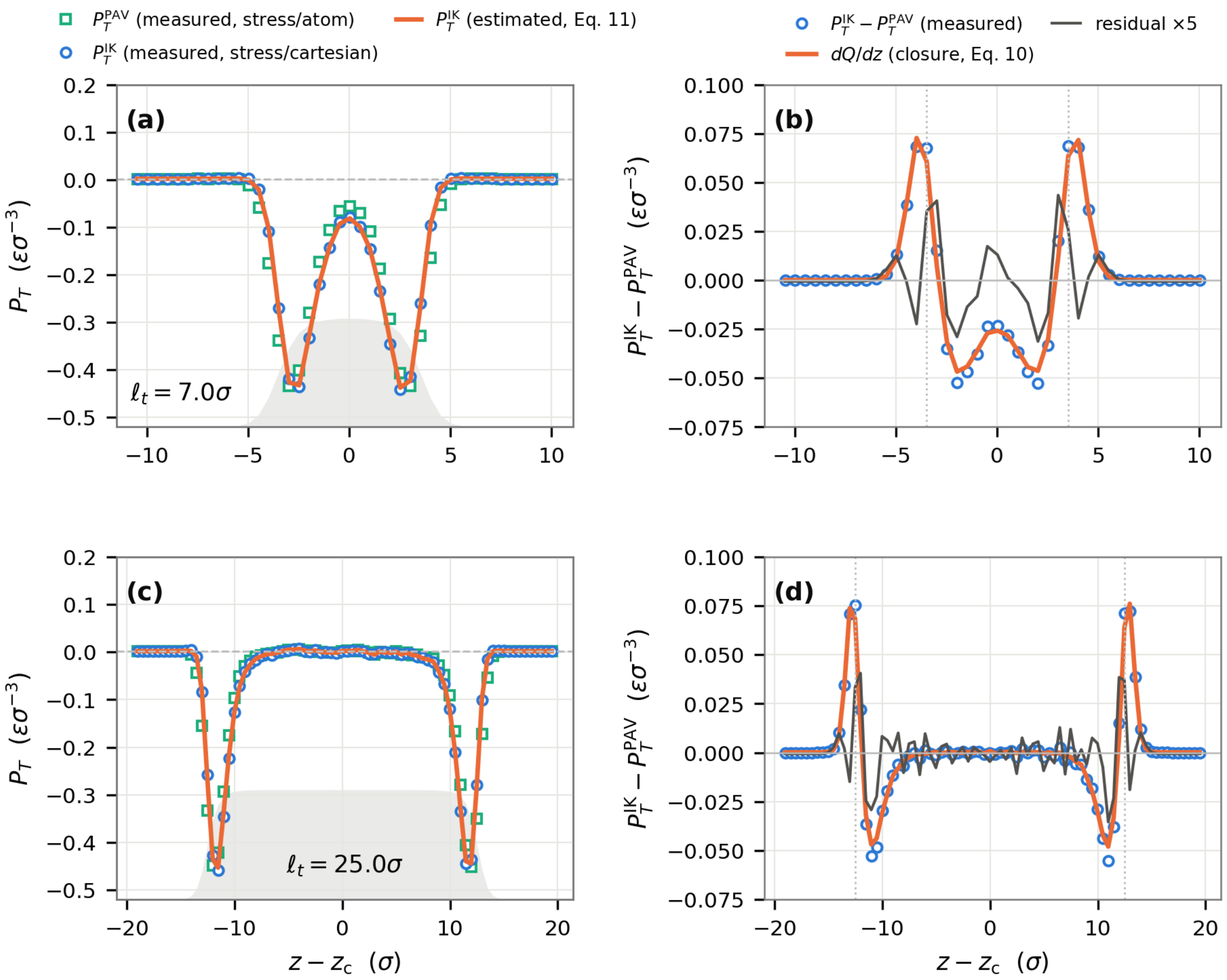

***Figure 2.*** *Tangential pressure profiles of the thinnest [(a), (b); $\ell_t = 7.0\sigma$] and thickest [(c), (d); $\ell_t = 25.0\sigma$] films at $T = 0.72$, $r_c = 7.5\sigma$. (a), (c): measured PAV (squares) and IK (circles) tangential profiles, and the IK profile estimated from the PAV one with Eq. (11) (line); the dashed line is the vapor pressure and the shaded area is the density profile (arbitrary scale). (b), (d): measured contour difference $P_T^{\rm IK} - P_T^{\rm PAV}$ (circles), closure prediction $\mathrm{d}Q/\mathrm{d}z$ from Eq. (10) with the pair-local kernel (line), and residual multiplied by five (thin line). Dotted verticals mark the half-density points.*

The residual (thin line, $\times 5$) has two components. In the liquid interior it is a noise-like signal of amplitude $\sim 0.002$. At the interfaces it is a systematic doublet of amplitude up to 0.007–0.011, located on the liquid flank of the density profile, with the same sign pattern as the difference itself: the closure underestimates the contour difference there by about 10%. We return to its origin in Sec. V.

## B. Thickness series

Figure 3 collects the seven films of the thickness series. As the film thins, the two interfacial doublets of $P_T^{\rm IK} - P_T^{\rm PAV}$ approach each other, overlap and finally merge into a single structure with a negative plateau across the whole interior ($\ell_t = 7.0$ and $7.75\sigma$). The closure follows this evolution without any adjustable parameter: the same kernel $K(\zeta)$, computed once from the bulk $g(r)$, predicts the shape of the merged structure in the thinnest films and the separated doublets in the thickest. The rms residual is 0.0020–0.0022 for all seven films, against 0.0200–0.0205 for the uncorrected difference, a tenfold reduction that is independent of thickness (Table 1).

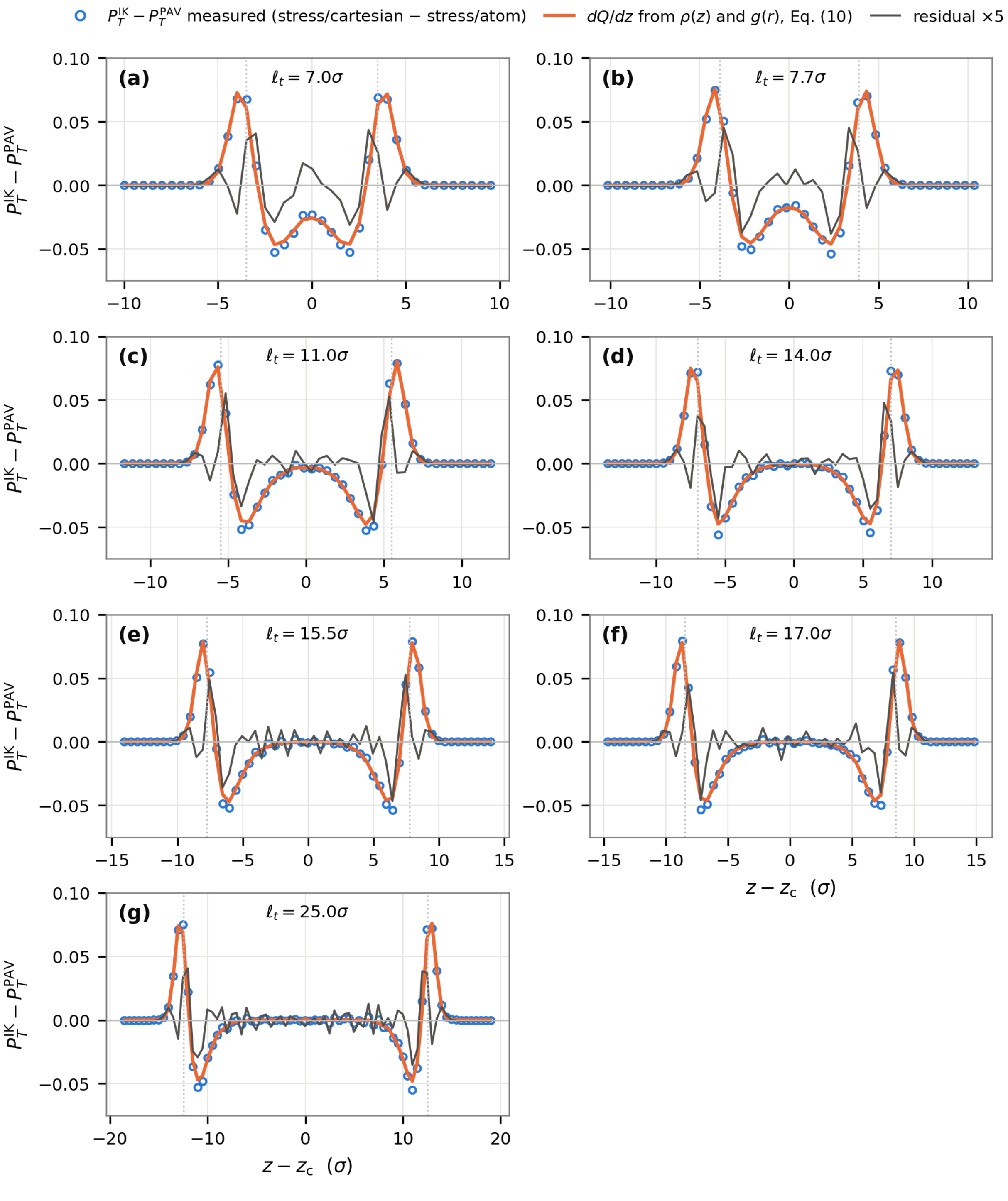

***Figure 3.*** *Thickness series at* $T = 0.72$, $r_c = 7.5\sigma$*: measured contour difference* $P_T^{\mathrm{IK}} - P_T^{\mathrm{PAV}}$ *(circles), closure prediction* $\mathrm{d}Q/\mathrm{d}z$ *(line) and residual multiplied by five (thin line) for films of* $\ell_t = 7.0$ *to* $25.0\sigma$ *[(a)–(g)]. Dotted verticals mark the half-density points. The rms of the residual over all bins is 0.0020–0.0022* $\epsilon\sigma^{-3}$ *for the seven films, against 0.0200–0.0205 for the uncorrected difference (Table 1).*

## C. Cutoff series

Changing the cutoff changes both the kernel and the system: $K(\zeta)$ has support $|\zeta| < r_c$, and the coexistence densities and the surface tension of the truncated fluid shift with $r_c$ ($\rho_\ell \simeq 0.81$, 0.82–0.83, 0.82–0.83 and $P_{\mathrm{vap}} = 0.0029,\ 0.0021,\ 0.0019$ for $r_c = 3.5,\ 5.0,\ 7.5\sigma$). Figure 4 shows the eight films at $r_c = 3.5$ and $5.0\sigma$. The interfacial doublets become narrower and smaller as $r_c$ decreases (amplitude $\approx 0.05$ at $r_c = 3.5$ versus 0.07 at 7.5), and the estimator follows: the rms residual is 0.0017–0.0021, against 0.014–0.019 uncorrected. The capture fraction at the interface, $1 - \mathrm{rms}_{\mathrm{res}}/\mathrm{rms}_{\mathrm{uncorr}}$, is 86–90% for all fifteen films at $T = 0.72$ irrespective of $r_c$.

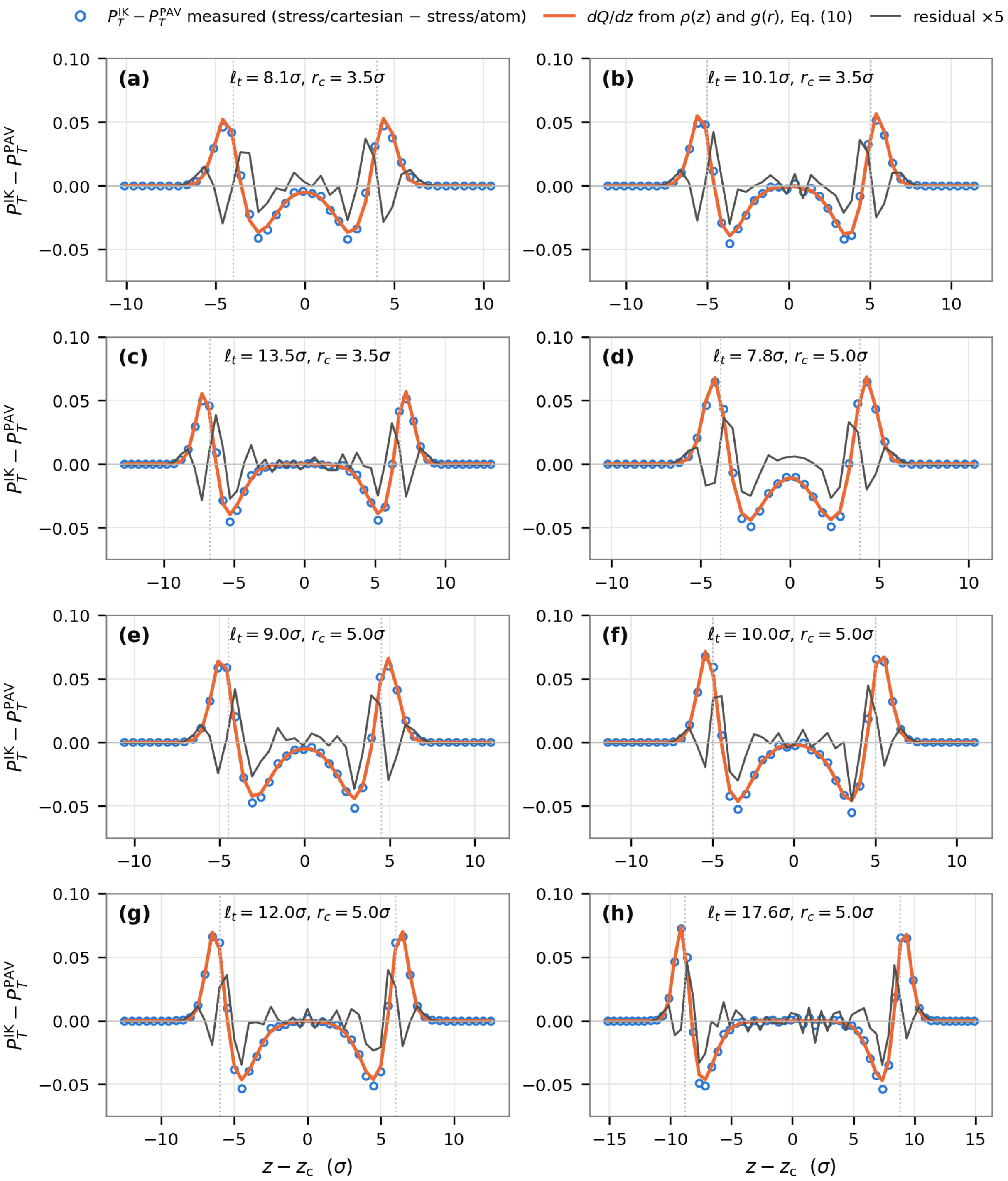

***Figure 4.*** *Cutoff series at $T = 0.72$: $r_c = 3.5\sigma$ [(a)–(c)] and $r_c = 5.0\sigma$ [(d)–(h)]. Symbols and lines as in Fig. 3; each series uses its own $g(r)$ and coexistence density. The rms of the residual over all bins is 0.0017–0.0018 at $r_c = 3.5\sigma$ and 0.0018–0.0021 at $r_c = 5.0\sigma$, against 0.0138–0.0147 and 0.0175–0.0190 for the uncorrected difference (Table 1).*

## D. Temperature

At $T = 0.90$ the interface is wider (10–90% width 3.1–3.5$\sigma$ against 2.2–2.5$\sigma$ at $T = 0.72$), the vapor is denser ($\rho_v = 0.0158$, $P_{\rm vap} = 0.0123$) and the liquid less structured ($\rho_\ell = 0.74$–$0.75$). All three changes favour the closure, and Fig. 5 shows that the estimator improves accordingly for the five films of the series: the rms residual drops to 0.0009–0.0012 (0.010–0.012 uncorrected), the interfacial capture rises to 87–93%, and in the liquid interior the residual is at the statistical noise ($\chi^2$ per bin 0.6–1.4).

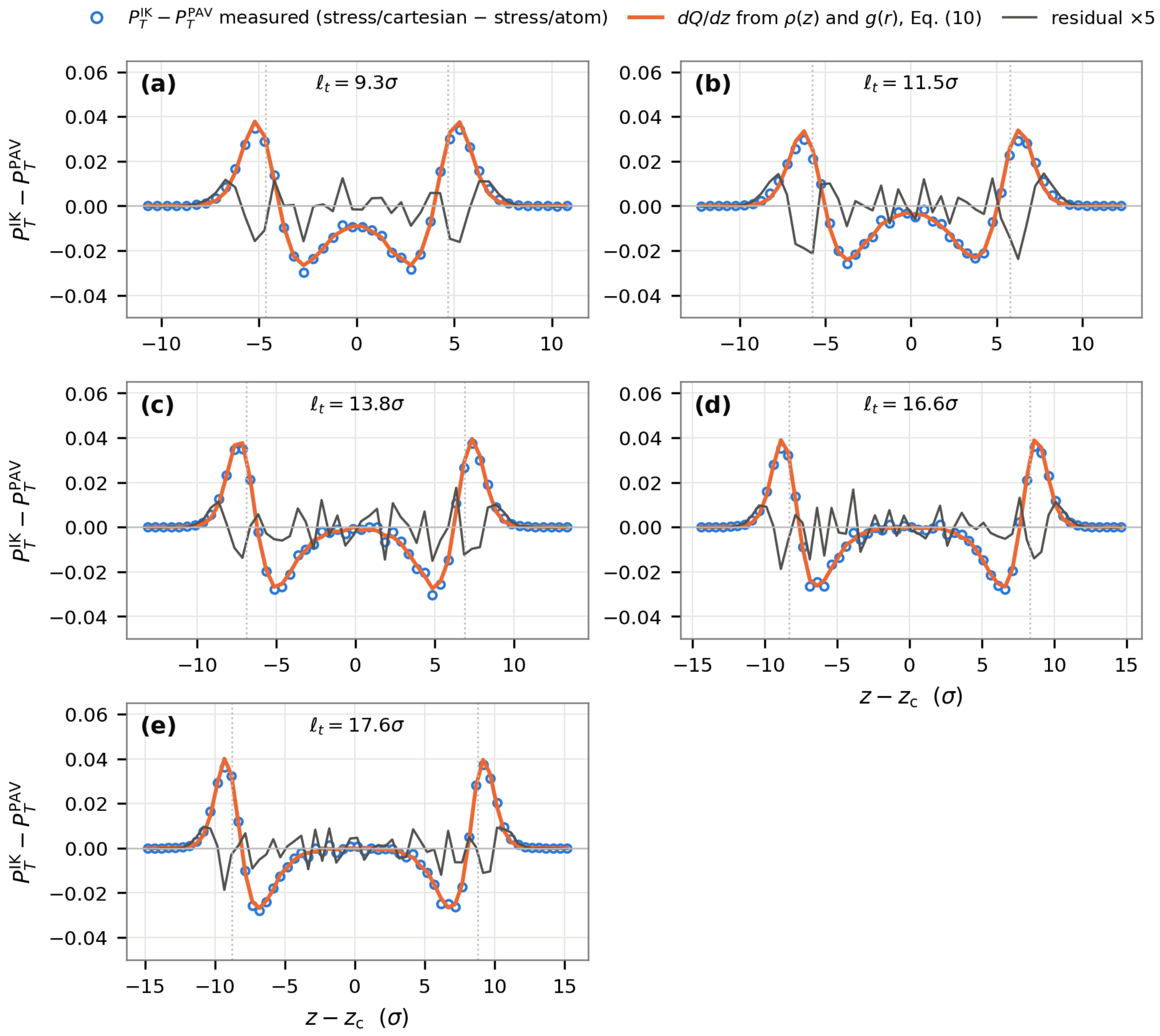

***Figure 5.*** *Temperature series at $T = 0.90$, $r_c = 7.5\sigma$: films of $\ell_t = 9.3$ to $17.6\sigma$ [(a)–(e)]. Symbols and lines as in Fig. 3; note the different vertical scale. The rms of the residual over all bins is 0.0009–0.0012, against 0.0100–0.0116 for the uncorrected difference (Table 1).*

## E. Central-plane anisotropy

Figure 6 summarizes the performance of the estimator over the 20 films. Panel (a) plots the measured difference against the predicted $\mathrm{d}Q/\mathrm{d}z$ bin by bin; the liquid-interior bins (coloured) fall on the diagonal over the whole range $-0.06$ to $+0.005$, while the interfacial bins (grey) scatter around it with the $\sim$ 10% deficit noted above. Panel (b) isolates the central plane, which is where a stability criterion is evaluated: the predicted and measured contour differences at $z_c$ agree to within 0.0012 over a range of 0.03, for all three cutoffs and both temperatures. Panel (c) shows the resulting anisotropy $\Delta P_T$ as a function of thickness. The IK value (filled symbols) decays from 0.104 at $\ell_t = 7\sigma$ to zero within noise beyond $\ell_t \approx 16\sigma$ at $r_c = 7.5\sigma$, and the estimated value (crosses) tracks it to within 0.0012 at every point, whereas the uncorrected PAV value (open symbols) underestimates the anisotropy by 22–28% in the thin films (0.076 against 0.104 at $7\sigma$). The PAV and IK values converge only when the central plane is more than $r_c$ away from both interfaces, i.e. when the film has a genuine bulk interior; it is in exactly the thin-film regime of interest that the contour correction is largest and the estimator is needed.

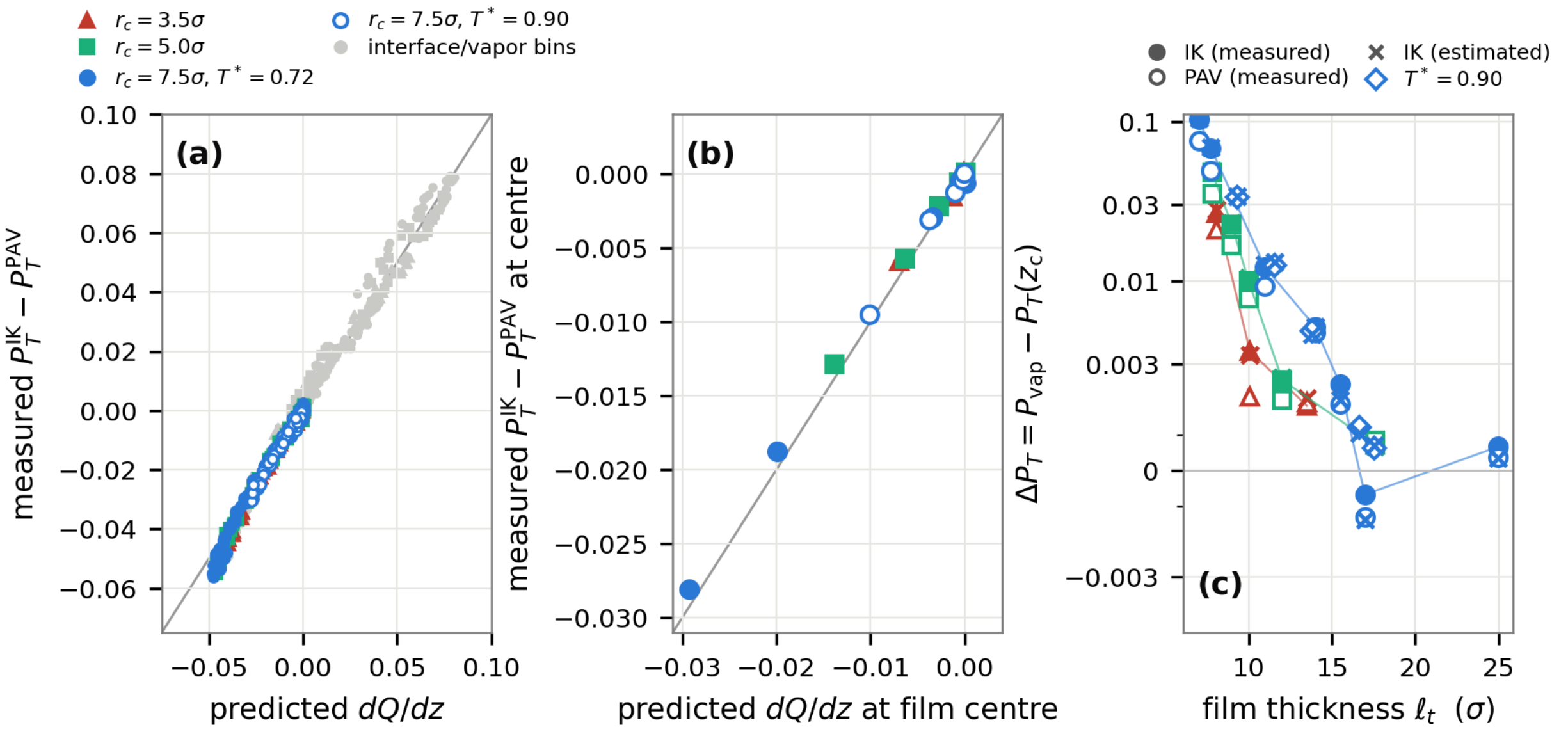

***Figure 6.*** *(a) Measured contour difference versus closure prediction, bin by bin, for the 20 films; liquid-interior bins ($\rho > 0.9\rho_{\max}$) are coloured by cutoff (filled: $T = 0.72$; open: $T = 0.90$), interfacial and vapor bins are grey. (b) The same comparison restricted to the central plane of each film ($|z - z_c| \leq 1\sigma$). (c) Central-plane anisotropy $\Delta P_T = P_{\rm vap} - P_T(z_c)$ versus thickness from the measured IK profile (filled), the measured PAV profile (open) and the estimated IK profile (crosses), for $r_c = 3.5$ (triangles), 5.0 (squares) and 7.5σ (circles) at $T = 0.72$, and for $T = 0.90$ (diamonds); the vertical scale is linear below $|\Delta P_T| = 0.003$ and logarithmic above.*

## F. Residual statistics

Figure 7 and Table 1 quantify the residual by region. In the liquid interior the rms residual is 0.0020–0.0037 at $T = 0.72$, to be compared with a standard error per bin of 0.0014–0.0015 for the measured difference: $\chi^2$ per bin is 2–7, largest for the thinnest films, where the "liquid" region is only a few bins wide and borders the interfacial deficit. At $T = 0.90$ it is 0.6–1.4, i.e. the residual is noise. Given that the residual is evaluated with a closure that has no adjustable parameters, we regard a liquid-interior residual within a factor $\lesssim 2$ of the statistical noise of a $2 \times 10^6$-step run as the practical limit of the method; longer runs would reduce the noise but not, presumably, the small systematic part. At the interface the residual rms is 0.0042–0.0063 at $T = 0.72$ and 0.0017–0.0026 at $T = 0.90$, against 0.033–0.055 and 0.021–0.025 uncorrected. In the vapor both are below $5 \times 10^{-4}$.

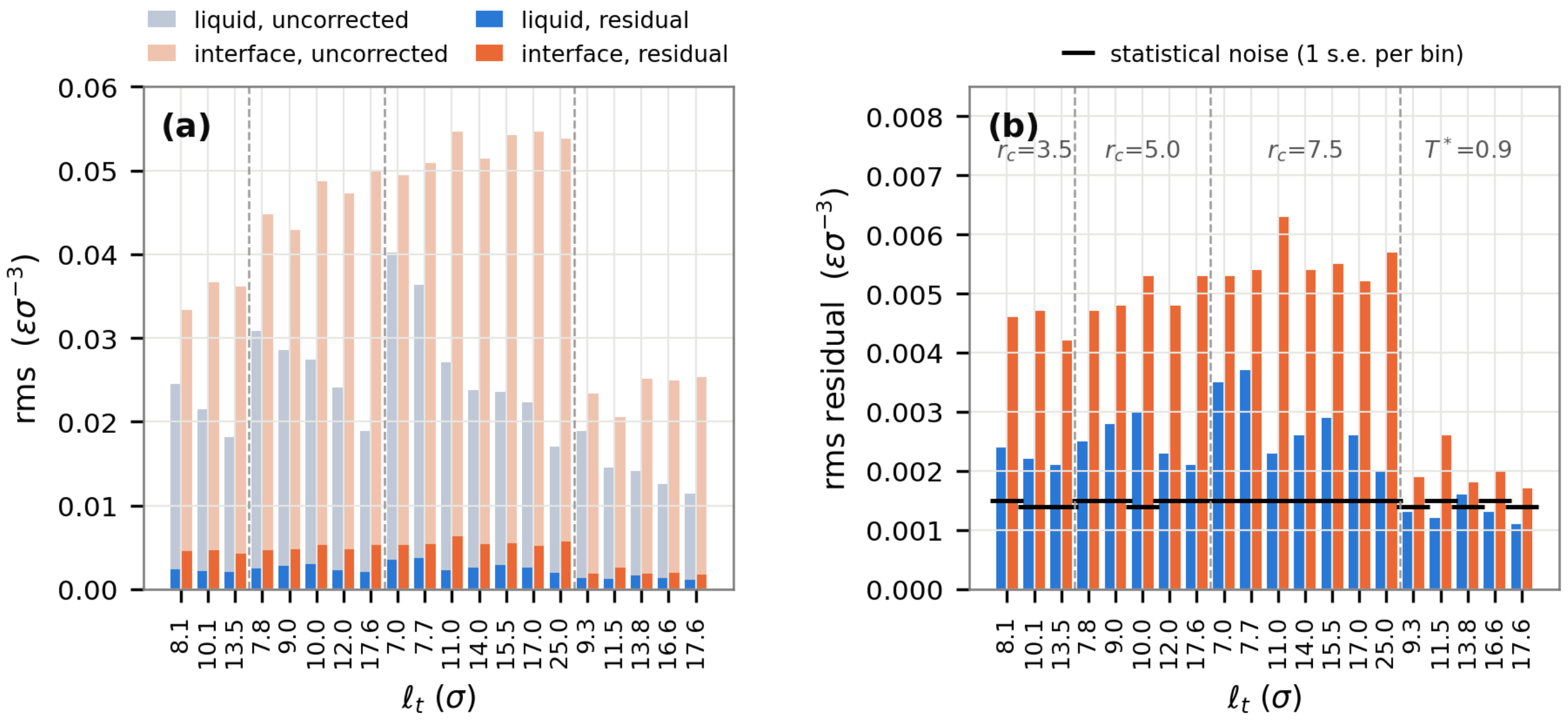

***Figure 7.*** *(a) rms of the uncorrected contour difference (light bars) and of the residual of the estimator (dark bars) in the liquid interior and at the interface for the 20 films, grouped by series. (b) rms residual on an expanded scale; horizontal marks are the standard error per bin of the measured difference in the liquid.*

***Table 1.*** *The 20 films: temperature, cutoff, thickness $\ell_t$ (distance between half-density points), number of atoms $N$, central density $\rho_c$, vapor pressure $P_{\mathrm{vap}}$ (IK normal pressure in the vapor), central-plane anisotropy $\Delta P_T$ from the measured IK and PAV profiles and from the estimate, rms over all bins of the uncorrected difference $P_T^{\mathrm{IK}} - P_T^{\mathrm{PAV}}$ and of the residual of the estimator, $\chi^2$ per bin of the residual in the liquid interior, and surface tension $\gamma = 1/2 \int \left(P_N^{\mathrm{IK}} - P_T\right) \mathrm{d}z$, which is identical to the digits shown whether $P_T^{\mathrm{IK}}$, $P_T^{\mathrm{PAV}}$ or $P_T^{\mathrm{IK,est}}$ is used (Sec. IV.G). All films: $b_t = 49.6\sigma$, $L_z = 48\sigma$, bins of $0.5\sigma$, 200 blocks of $10^4$ steps. Pressures in $\epsilon\sigma^{-3}$, $\gamma$ in $\epsilon\sigma^{-2}$, lengths in $\sigma$.*

| $T$ | $r_c$ | $\ell_t$ | $N$ | $\rho_c$ | $P_{\mathrm{vap}}$ | $\Delta P_T^{\mathrm{IK}}$ | $\Delta P_T^{\mathrm{PAV}}$ | $\Delta P_T^{\mathrm{est}}$ | $\mathrm{rms}_{\mathrm{uncorr}}$ | $\mathrm{rms}_{\mathrm{res}}$ | $\chi^2_{\mathrm{liq}}$ | $\gamma$ |
|---|---|---|---|---|---|---|---|---|---|---|---|---|
| 0.72 | 3.5 | 8.07 | 16396 | 0.8118 | 0.0029 | +0.0268 | +0.0209 | +0.0279 | 0.0138 | 0.0017 | 2.9 | 0.8317 |
| 0.72 | 3.5 | 10.08 | 20420 | 0.8140 | 0.0029 | +0.0036 | +0.0021 | +0.0034 | 0.0147 | 0.0018 | 2.9 | 0.8186 |
| 0.72 | 3.5 | 13.49 | 27209 | 0.8144 | 0.0029 | +0.0018 | +0.0019 | +0.0020 | 0.0147 | 0.0017 | 2.5 | 0.8342 |
| 0.72 | 5.0 | 7.80 | 15942 | 0.8232 | 0.0021 | +0.0481 | +0.0353 | +0.0491 | 0.0178 | 0.0018 | 2.8 | 0.9725 |
| 0.72 | 5.0 | 8.99 | 18378 | 0.8257 | 0.0021 | +0.0225 | +0.0168 | +0.0231 | 0.0175 | 0.0019 | 3.8 | 0.9748 |
| 0.72 | 5.0 | 10.01 | 20420 | 0.8268 | 0.0021 | +0.0099 | +0.0077 | +0.0105 | 0.0188 | 0.0021 | 4.7 | 0.9686 |
| 0.72 | 5.0 | 12.02 | 24504 | 0.8274 | 0.0021 | +0.0026 | +0.0020 | +0.0026 | 0.0187 | 0.0019 | 2.7 | 0.9882 |
| 0.72 | 5.0 | 17.62 | 35869 | 0.8276 | 0.0021 | +0.0007 | +0.0008 | +0.0007 | 0.0190 | 0.0020 | 2.3 | 0.9822 |
| 0.72 | 7.5 | 7.00 | 14294 | 0.8221 | 0.0019 | +0.1042 | +0.0761 | +0.1054 | 0.0202 | 0.0020 | 5.7 | 1.0470 |
| 0.72 | 7.5 | 7.74 | 15835 | 0.8263 | 0.0019 | +0.0679 | +0.0492 | +0.0690 | 0.0200 | 0.0021 | 6.6 | 1.0556 |
| 0.72 | 7.5 | 10.98 | 22461 | 0.8314 | 0.0019 | +0.0122 | +0.0093 | +0.0126 | 0.0205 | 0.0022 | 2.7 | 1.0543 |
| 0.72 | 7.5 | 13.98 | 28587 | 0.8320 | 0.0019 | +0.0052 | +0.0047 | +0.0052 | 0.0203 | 0.0022 | 3.0 | 1.0481 |
| 0.72 | 7.5 | 15.48 | 31651 | 0.8322 | 0.0019 | +0.0024 | +0.0019 | +0.0020 | 0.0203 | 0.0022 | 4.5 | 1.0504 |
| 0.72 | 7.5 | 16.98 | 34713 | 0.8321 | 0.0019 | -0.0007 | -0.0013 | -0.0014 | 0.0203 | 0.0021 | 3.3 | 1.0487 |
| 0.72 | 7.5 | 24.99 | 51048 | 0.8320 | 0.0019 | +0.0007 | +0.0004 | +0.0004 | 0.0205 | 0.0022 | 2.0 | 1.0526 |
| 0.90 | 7.5 | 9.34 | 18451 | 0.7440 | 0.0123 | +0.0334 | +0.0240 | +0.0341 | 0.0112 | 0.0009 | 0.9 | 0.6528 |
| 0.90 | 7.5 | 11.54 | 22461 | 0.7480 | 0.0124 | +0.0125 | +0.0094 | +0.0132 | 0.0100 | 0.0012 | 0.7 | 0.6642 |
| 0.90 | 7.5 | 13.80 | 26546 | 0.7494 | 0.0123 | +0.0049 | +0.0036 | +0.0046 | 0.0115 | 0.0011 | 1.4 | 0.6599 |
| 0.90 | 7.5 | 16.62 | 31651 | 0.7499 | 0.0123 | +0.0012 | +0.0008 | +0.0011 | 0.0114 | 0.0011 | 0.8 | 0.6578 |
| 0.90 | 7.5 | 17.55 | 33312 | 0.7497 | 0.0124 | +0.0006 | +0.0007 | +0.0007 | 0.0116 | 0.0009 | 0.6 | 0.6663 |

## G. Contour invariance of the surface tension

The last column of Table 1 gives $\gamma = 1/2 \int \left(P_N^{\mathrm{IK}} - P_T\right) \mathrm{d}z$ for each film. The value is the same, to all four digits shown and for every one of the 200 blocks of every film, whether $P_T^{\mathrm{IK}}$, $P_T^{\mathrm{PAV}}$ or the estimated profile is used, and also when $P_N^{\mathrm{PAV}}$ is used in place of $P_N^{\mathrm{IK}}$. This is not a statistical agreement but an identity: both weights are normalized, so the integral over $z$ of each pair's contribution is its full virial in either case, and the two profiles have the same integral configuration by configuration when they are sampled on the same time steps, as they are here. (If the two profiles are accumulated on different time steps the integrals differ by the statistical fluctuation of the virial between the two samples; we observed this in a short test run before matching the sampling of the two computes.) The interfacial doublets of Figs. 3 and 4, whose amplitude is $\pm 0.05$–$0.08$, thus integrate to zero exactly, which is the content of Eq. (5). The estimator inherits the property because the discrete $\mathrm{d}Q/\mathrm{d}z$ of Eq. (10) integrates to zero on the data grid: the correction redistributes the tangential stress between the interface and the interior without changing its integral. For the symmetric films studied here the same holds for each half of the cell, because $Q(z_c) = 0$ by symmetry; it is only for non-equivalent interfaces that the half-cell integrals become contour dependent, by $\pm Q(z_s)$ (Sec. II.A), and that the estimated $Q(z)$ of Eq. (10) acquires a direct use. The tensions themselves depend on the cutoff ($\gamma = 0.82$–$0.83$, $0.97$–$0.99$ and $1.05$–$1.06$ for $r_c = 3.5$, $5.0$ and $7.5\sigma$ at $T = 0.72$) and on temperature ($0.65$–$0.67$ at $T = 0.90$), as expected for the truncated potential [30,49,50]. They are consistent with the literature once the truncation scheme is matched: LAMMPS integrates the truncated force and computes the virial from it, which is the spherically truncated and shifted (STS) model in the terminology of Ref. [30] (the same distinction matters for the test-area and volume-perturbation routes to $\gamma$ [51–53]), for which that work quotes $\gamma = 0.91$ at $T = 0.72$, $r_c = 4.4\sigma$, $0.616 \pm 0.003$ at $T = 0.92$, $r_c = 5.5\sigma$, and $0.63 \pm 0.02$ at $T = 0.92$, $r_c = 7.33\sigma$; linear interpolation of our $r_c = 3.5$ and $5.0\sigma$ values gives $0.92$ at $4.4\sigma$, and our $r_c = 7.5\sigma$ tensions at $T = 0.90$ correspond to $0.61$–$0.62$ at $T = 0.92$ with the temperature slope between our two isotherms, within the uncertainty of the $7.33\sigma$ value. (The spherically truncated model with the impulsive force at $r_c$ included in the virial gives higher values, 0.992 and 1.042 at $T = 0.72$ for $r_c = 4.4$ and $5.5\sigma$ [30], and should not be compared with LAMMPS `lj/cut` results.) The variation within a series (up to 0.02) exceeds the block standard error of a single $2 \times 10^6$-step run ($\approx 0.002$) and reflects the dependence of the tension of a thin film on its thickness [29], which is not the subject of this paper.

## H. Normal component: flatness test

Figure 8 shows the test of Sec. II.F for a thin and a thick film at $T = 0.72$ and for a thin film at $T = 0.90$. The $zz$ component of the PAV profile is strongly structured across each interface, with excursions of $-0.19$ to $+0.14$ about $P_{\mathrm{vap}}$ (rms 0.038–0.053 over the cell at $T = 0.72$, 0.027–0.030 at $T = 0.90$), in a region where the IK normal pressure is flat to within 0.001–0.002. Adding $\mathrm{d}Q_N/\mathrm{d}z$ from the closure with the normal kernel (17), computed from the same density profile and the same $g(r)$ as the tangential correction, removes 83–87% of this structure at $T = 0.72$ (rms 0.0060–0.0073, maximum 0.030–0.038) and 88–92% at $T = 0.90$ (rms 0.0024–0.0033, maximum 0.008–0.013), for all 20 films. The residual has the same character as the tangential one: a doublet on the liquid flank of each interface, and nothing in the interior or the vapor. The test uses only the per-atom stress and the density; it requires neither the IK run nor any assumption about the interface, since the answer, a constant, is fixed by mechanical equilibrium, and it can therefore be applied to any PAV run to gauge the closure error before trusting the tangential estimate.

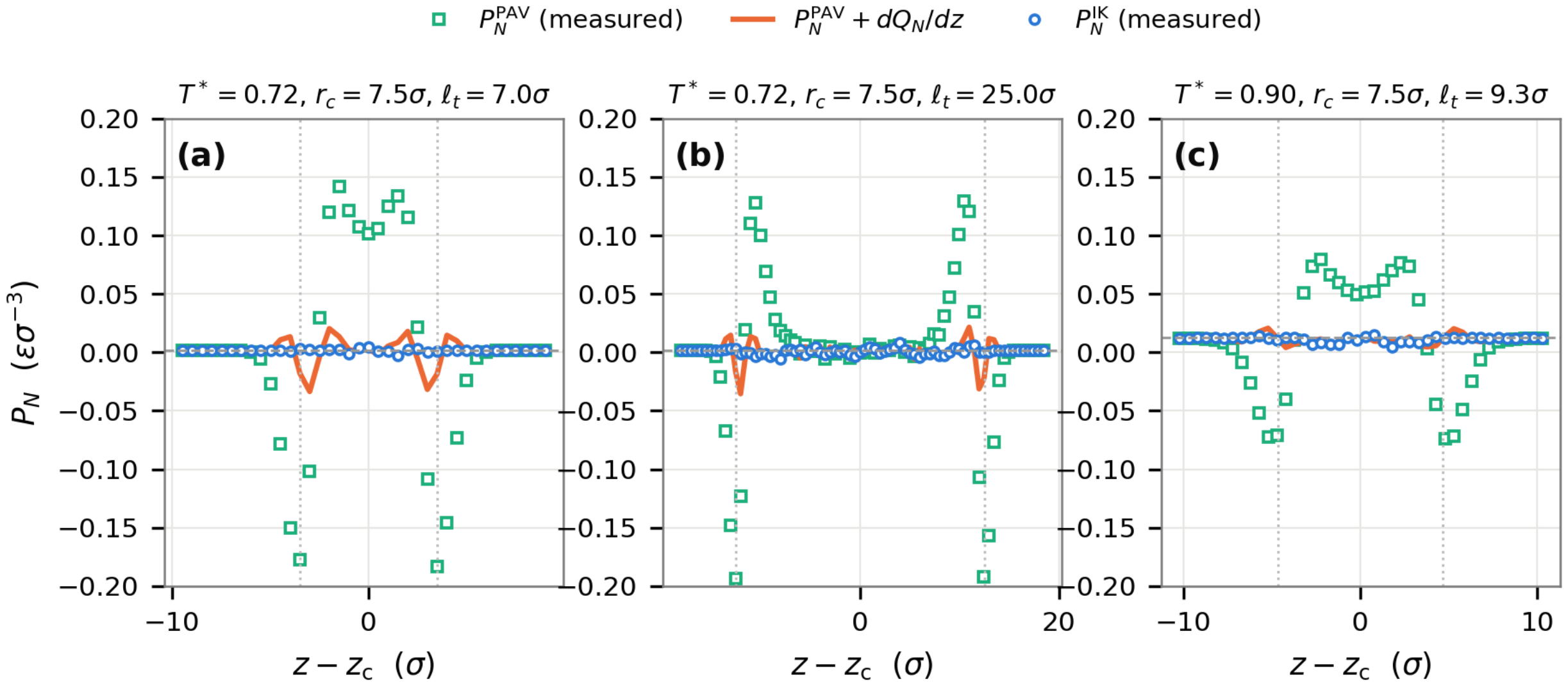

***Figure 8.*** *Flatness test of the normal component for (a) the* $\ell_t = 7.0\sigma$ *and (b) the* $25.0\sigma$ *films at* $T = 0.72$, $r_c = 7.5\sigma$, *and (c) the* $9.3\sigma$ *film at* $T = 0.90$*: measured zz component of the PAV profile (squares), the same corrected with* $\mathrm{d}Q_N/\mathrm{d}z$ *from Eqs. (10) and (17) (line), and measured IK normal pressure (circles). The dashed line is* $P_{\text{vap}}$*; dotted verticals mark the half-density points. The rms over all bins of the corrected profile about* $P_{\text{vap}}$ *is 0.0069, 0.0073 and 0.0024* $\epsilon\sigma^{-3}$ *in (a), (b) and (c), against 0.053, 0.052 and 0.029 for the uncorrected one (Table S2 of the supplementary material).*

## I. Accuracy of the gauge term itself

For the partition of the tension between two interfaces it is $Q(z)$, not its derivative, that matters (Sec. II.A). Figure 9 compares $Q$ obtained by integrating the measured difference $P_T^{\text{IK}} - P_T^{\text{PAV}}$ with the closure estimate of Eq. (10), both referred to zero in the vapor. $Q$ is an odd function about the film centre with extrema of $\mp 0.07$ to $0.11\,\epsilon\sigma^{-2}$ located near the half-density points, growing with the cutoff (0.07, 0.09–0.10 and 0.10–0.11 for $r_c = 3.5$, 5.0 and $7.5\sigma$) and smaller at $T = 0.90$ (0.06–0.07); at $r_c = 7.5\sigma$ its maximum is 10% of $\gamma$. This is the amount of tension that the contour choice would move from one interface to the other if a dividing plane were placed at the position of the extremum, and it does not decrease with thickness, because it is set by the pairs that straddle each interface. The closure reproduces $Q$ with a maximum error inside the film ($|z - z_c| < \ell_t/2$) of 0.003–0.006 at $T = 0.72$ and 0.001–0.002 at $T = 0.90$, i.e. 4–6% and 1–4% of the extremum and below 0.6% of $\gamma$, for all 20 films (Table S3 of the supplementary material); the error is largest on the liquid flank of each interface, where the $\mathrm{d}Q/\mathrm{d}z$ residual of Figs. 3–5 is concentrated, and it does not accumulate across the interior, so that $Q$ at any plane more than $2\sigma$ inside the half-density points is recovered to within 0.002, i.e. 0.2% of $\gamma$.

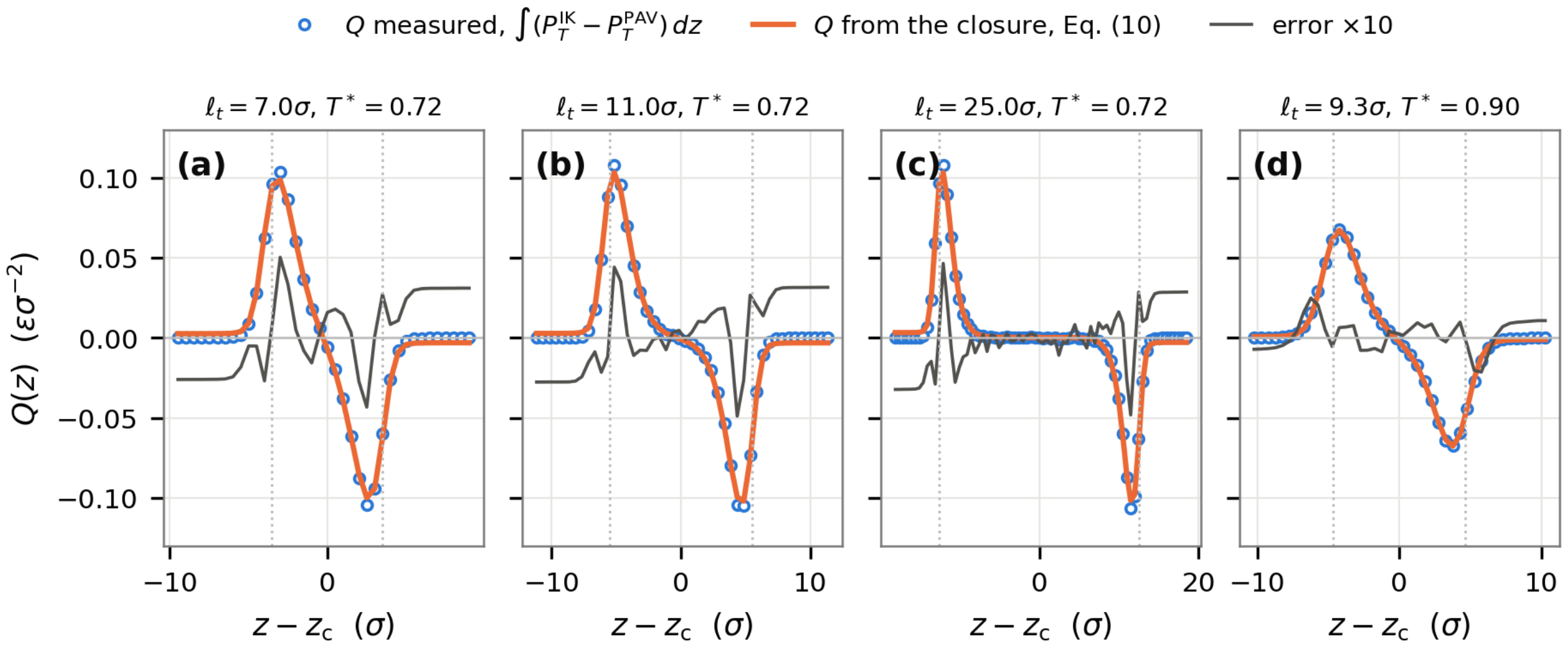

***Figure 9.*** *The gauge term $Q(z)$ for films of (a) $\ell_t = 7.0$, (b) $11.0$ and (c) $25.0\sigma$ at $T = 0.72$, $r_c = 7.5\sigma$, and (d) $9.3\sigma$ at $T = 0.90$: integral of the measured difference $P_T^{\mathrm{IK}} - P_T^{\mathrm{PAV}}$ (circles), closure estimate from Eq. (10) (line), and their difference multiplied by ten (thin line); all referred to zero in the vapor. Dotted verticals mark the half-density points.*

### J. Films with non-equivalent interfaces

Figure 10 shows film III of Sec. III.A, the one with a strictly clean lower face and a stationary adsorbed layer ($\Gamma_2 = 0.193\sigma^{-2}$, 60% of a monolayer of the large adsorbate); films I and II are shown in the supplementary material, and Table 2 collects the results of the three. The B density peaks sit 0.3–$1\sigma$ outside the half-density points of A, B is practically absent from the interior, and the film is thinner than $2r_c$, so no plane inside it is bulk-like. The tangential profile of the decorated face [Fig. 10(b)] is visibly different from that of the clean face, shallower by 20% (by 40% in film II) and wider. The mixture form of the closure (Sec. II.D) reproduces the measured IK profile with an accuracy comparable to that in the pure films: rms residual 0.0019–0.0023 over all bins (uncorrected 0.017–0.018), 0.0023–0.0027 in the liquid and 0.0050–0.0056 at the interfaces, with the largest residual, 0.008–0.011, on the liquid flank of the decorated face; the gauge term, whose extrema are $\pm 0.10$, is recovered to within 0.004–0.007, and the corrected normal profile $P_N^{\mathrm{PAV}} + \mathrm{d}Q_N/\mathrm{d}z$ reproduces the measured $P_N^{\mathrm{IK}}$ with an rms of 0.007–0.008 against 0.043–0.046 for the uncorrected one, i.e. 82–85% of the structure is removed. The films with the large adsorbate are at the upper edge of the error range of Table 1, as expected for pairs for which the conformal $g$ is a cruder approximation; they are still within it. Figure 10(e) shows this test for film III over the whole cell: the spurious structure of $P_N^{\mathrm{PAV}}$ at the two faces, which reaches $\mp 0.17$ at the clean face and $\mp 0.10$ at the decorated one, is reduced to $\pm 0.03$ by the closure, with the same residual pattern on the liquid flank of each face as in the pure films (Fig. 8), and the measured $P_N^{\mathrm{IK}}$ is flat on the scale of the figure. On a finer scale $P_N^{\mathrm{IK}}$ is $0.0019$ in the lower vapor and $0.0052$ in the upper one (Sec. III.A): the difference, the partial pressure of the B vapor, is sustained by the wall on B at the far end of the upper gap and transmitted to the film by the momentum-conserving fix, which balances the wall force by a uniform body force on all atoms, so that $P_N^{\mathrm{IK}}$ changes by this amount across the film, where the mass is, and falls back to $0.0019$ inside the wall region $z > 44\sigma$, from which B is excluded. This gradient, of order $10^{-3}$ over $10\sigma$, is irrelevant to the flatness test. The per-interface tensions are evaluated with the local $P_N^{\mathrm{IK}}(z)$, which carries it, and in the liquid the tangential pressure carries the same hydrostatic gradient, so that the anisotropy, and with it $\gamma_k$, is not affected to first order; replacing $P_N$ by the vapor pressure of each side would instead shift $\gamma_2$ of film III by 0.024, because it would assign a spurious anisotropy to the wall region. The surface tension is again identical for the three tangential profiles, $\gamma = 1/2\,(\gamma_1 + \gamma_2)$ with $\gamma_1$ and $\gamma_2$ the integrals of $P_N - P_T$ over the two halves of the cell, split at the bin edge $z_s$ nearest the film centre. The two faces are clearly resolved. In film III the clean face has $\gamma_1 = 1.049$ with IK, which is the tension of the pure films at this cutoff, $\gamma_0 = 1.051 \pm 0.003$ (Table 1), while the decorated face has $\gamma_2 = 0.859$; in film II, $\gamma_1 = 1.021$ and $\gamma_2 = 0.658$. The block standard error is about 0.003 on each, and the estimator reproduces the IK partition to 0.0003 in all three films. The difference between the IK and PAV partitions, $-0.0013$ to $-0.0014$, is $-Q(z_s)$ to within 0.0002, as Eq. (5) requires, and the closure gives it to within 0.0004. It is small in all three films because the asymmetry of the density within $r_c$ of the centre is small, the two faces differing only by a thin adsorbed layer: $Q$ is large at each face ($\pm 0.10$) and nearly cancels at the midplane, which is also why it would not be small for a thinner film, a supported film, or a dividing plane placed off-centre.

The per-interface tensions have a simple physical reading. That the clean face of film III recovers $\gamma_0$ to within the statistical error, with a dividing plane only $5\sigma$ from either interface, is the most direct evidence that the partition at a plane inside a film with no bulk interior is physically meaningful once the contour is fixed. For the decorated faces, the surface pressures of the adsorbate, $\Pi_k = \gamma_0 - \gamma_k$, give $\Pi/(\Gamma k_B T) = 1.0$–$1.2$ for the two faces of film I (the range covering whether or not the B in the vapor is counted in $\Gamma$): the small adsorbate behaves as a nearly ideal two-dimensional gas at these coverages. For the large adsorbate the ratio is 1.4 at $\Gamma = 0.03\sigma^{-2}$ (film II, lower face, where $\Pi_1 = 0.030$ is only ten times its uncertainty), 1.4 at $0.19\sigma^{-2}$ (film III) and 2.4–2.7 at $0.21\sigma^{-2}$ with the stronger A–B attraction (film II), far from the ideal limit. These checks do not depend on the method; what the method adds is that the contour transfer, 0.15% of $\gamma$, is 2.5–4% of the surface pressure of the dilute faces of films I and II and would be the dominant uncertainty of the surface pressure of a dilute adsorbate on a thinner film. The drift of the coverage during the runs of films I and II (Sec. III.A) is visible in the halves of the production: $\gamma_1 = 0.998$ and $0.992$ in film I, 1.028 and 1.015 in film II, while $\gamma_2$ changes by less than 0.006; in film III the two halves give $\gamma_1 = 1.050, 1.054$ and $\gamma_2 = 0.857, 0.855$.

***Table 2.*** *The three two-component films: adsorbate parameters, adsorbed coverages of the lower (1) and upper (2) faces, fraction of B in the vapor, rms residual of the estimator over all bins (uncorrected rms in parentheses) and at the interfaces, maximum error of the closure estimate of $Q$ inside the film, rms of $P_N^{\mathrm{PAV}} - P_N^{\mathrm{IK}}$ before and after correction, surface tension, per-interface tensions with the IK, PAV and estimated profiles, the measured and closure values of $-Q(z_s)$, and the reduced surface pressure of the adsorbate on each face. Pressures in $\epsilon\sigma^{-3}$, tensions and $Q$ in $\epsilon\sigma^{-2}$, coverages in $\sigma^{-2}$.*

| | Film I | Film II | Film III |
|---|---|---|---|
| $\sigma_{BB}, \sigma_{AB}, \epsilon_{AB}$ | 1.5, 1.25, 0.7 | 2.0, 1.5, 0.7 | 2.0, 1.5, 0.5 |
| $\Gamma_1, \Gamma_2$ | 0.071, 0.159 | 0.029, 0.205 | 0, 0.193 |
| B in the vapor | 9% | 8% | 24% |
| rms residual (uncorr.) | 0.0019 (0.018) | 0.0023 (0.017) | 0.0021 (0.017) |
| rms residual, interfaces | 0.0050 | 0.0056 | 0.0054 |
| max error of $Q$ | 0.004 | 0.007 | 0.005 |
| $P_N$ rms, before/after | 0.046/0.007 | 0.045/0.008 | 0.043/0.008 |
| $\gamma$ | 0.9528(20) | 0.8397(22) | 0.9538(20) |
| $\gamma_1, \gamma_2$ (IK) | 0.9952, 0.9104 | 1.0213, 0.6580 | 1.0486, 0.8590 |
| $\gamma_1, \gamma_2$ (PAV) | 0.9966, 0.9090 | 1.0226, 0.6567 | 1.0498, 0.8577 |
| $\gamma_1, \gamma_2$ (est.) | 0.9953, 0.9104 | 1.0210, 0.6583 | 1.0484, 0.8592 |
| $-Q(z_s)$, meas./closure | −0.0012/−0.0012 | −0.0012/−0.0016 | −0.0015/−0.0017 |
| $\Pi_k/\Gamma_k k_B T$, faces 1, 2 | 1.1, 1.2 | 1.4, 2.7 | –, 1.4 |

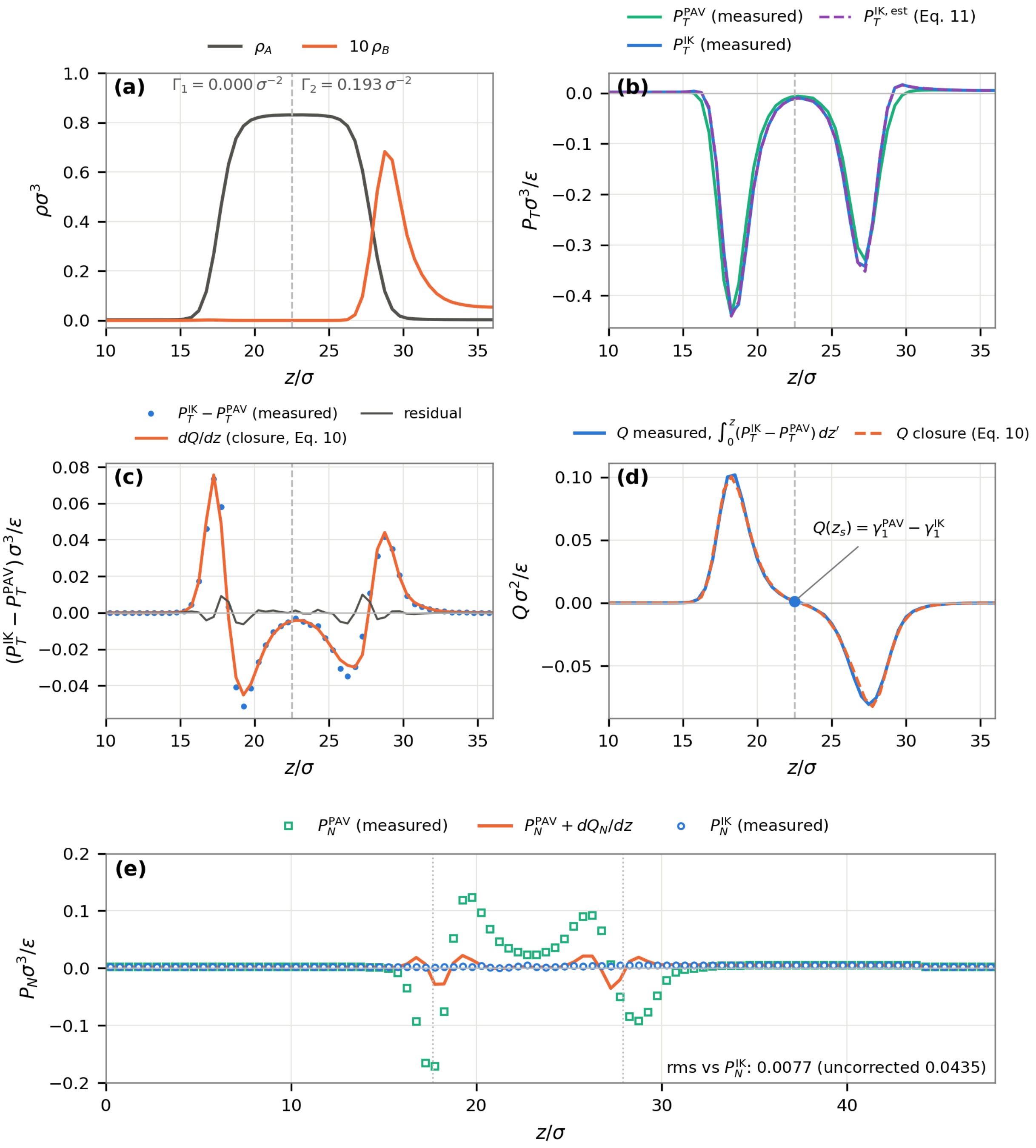

***Figure 10.*** *Two-component film III with non-equivalent interfaces ($\ell_t = 10.2\sigma$, $T = 0.72$, $r_c = 7.5\sigma$; Sec. III.A). (a) Density profiles of the liquid A and of the adsorbate B (the latter multiplied by ten), with the adsorbed coverage of each face; the dashed vertical is the dividing plane $z_s$. (b) Tangential pressure profiles: measured PAV and IK, and the estimate of Eq. (11) from the mixture closure. (c) Measured difference $P_T^{\mathrm{IK}} - P_T^{\mathrm{PAV}}$ (circles), closure $\mathrm{d}Q/\mathrm{d}z$ (line) and residual (thin line). (d) Gauge term $Q(z)$: integral of the measured difference and closure estimate; its value at the dividing plane is the amount of tension that the contour choice moves from one face to the other. (e) Normal component over the whole cell: measured per-atom zz profile (squares), the same corrected with $\mathrm{d}Q_N/\mathrm{d}z$ from Eqs. (10) and (17) (line), and measured IK normal pressure (circles); dotted verticals mark the half-density points of A. The rms of the uncorrected and corrected profiles about $P_N^{\mathrm{IK}}$ is given in the panel.*

## V. DISCUSSION

*Why the closure works for $Q$ and not for $P_T$.* The absolute tangential pressure at a liquid–vapor interface is the small difference between a large positive kinetic-plus-repulsive contribution and a large negative attractive one, and its interfacial minimum arises from the anisotropy of the pair correlations across the interface, which $\rho(z_1)\rho(z_2)g(r_{12})$ cannot describe [2]. The contour difference, on the other hand, is a geometric reweighting of the same pair virials: $q(z; z_i, z_j)$ measures how far the plane $z$ sits from the midpoint of the pair, and $Q(z)$ is non-zero only where pairs cross $z$ asymmetrically, i.e. where the density differs on the two sides of $z$ within $r_c$. The dominant contribution to (10) comes from pairs with $|\zeta|$ of the order of $r_c$, for which $g \approx 1$ and the kernel is insensitive to the details of the liquid structure. This is why a single bulk $g(r)$ suffices, why the result is essentially insensitive to the precise $\rho_\ell$ used in the bulk run, and why even the step function $g = \Theta(r - \sigma)$ already removes two thirds of the rms contour difference (from 0.020 to 0.0065 for the $r_c = 7.5\sigma$ films). The same argument explains the limit of the method: the $\sim 10\%$ deficit at the liquid flank of the interface comes from the short pairs ($|\zeta| \lesssim 2\sigma$) that straddle the steep part of the density profile, for which the local $g$ is neither the liquid nor the dilute one. A density-dependent $g(r; \rho)$ interpolated through intermediate densities, or a more sophisticated closure, would presumably reduce this deficit; we have not pursued it because the residual is already below the level at which the contour choice affects any conclusion we are aware of.

*Temperature.* At $T = 0.90$ the closure performs better in every region. The interface is wider, so the density varies less over the range of the short pairs, and the liquid is less structured, so $g(r)$ deviates less from its mean-field form. We expect the method to be at its best near the critical region and at its worst near the triple point, where $T = 0.72$ is already close.

*Which runs need the IK pass.* For a study that sweeps thickness, cutoff and temperature over tens of films, the estimator reduces the cost of the sampling stage to that of a PAV run plus one bulk $g(r)$ run per $(T, r_c)$: a factor of 2.2 in wall time for the runs of this work (Sec. I), in which the pair forces run on a GPU and the IK compute, which is not GPU-accelerated and evaluates every pair on the CPU, is the only part that does not scale. The saving depends on how often the IK profile is sampled; with a sampling interval of $10^3$ steps instead of 200 it would be a factor of about 1.2, at the price of a noisier IK profile; it is largest precisely in the setting of this work, GPU-accelerated forces with a CPU-bound IK compute, and smaller when the forces themselves run on the CPU. We recommend keeping the IK pass (or the method of planes, for $P_N$) on one reference film per series as a check, and on any system whose density profile is qualitatively different from a planar film (droplets [4,5], cylinders and fluids in cylindrical pores [7,54], films with a dense adsorbed phase), for which the closure has not been tested. The pair form of $Q$ holds for any pair potential, and (7) can be generalized to multi-site molecules by applying the closure per site pair; whether the accuracy carries over to molecular liquids, where the IK–PAV difference includes intramolecular contributions, is an open question.

*Non-equivalent interfaces.* The identity of the tensions in Table 1 should not be read as making the contour question moot. It is the total tension that is invariant; the tension of each interface of a film with two different surfaces is not, when the film is thinner than about $2r_c$, and the amount transferred between the two by the contour choice is $Q(z_s)$ at the dividing plane, which is not small: $|Q|$ reaches 0.07–0.11 $\epsilon\sigma^{-2}$ inside the film, i.e. up to 10% of $\gamma$ (Sec. IV.I and Fig. 9), and the closure estimates it to within 0.003–0.006 at $T = 0.72$ and 0.002 at $T = 0.90$, below 0.6% of $\gamma$. In such systems, which include films with an adsorbed layer on one side, supported films, and free-standing films of mixtures with preferential segregation, the IK split is the physically motivated one, because the IK profile is the one in which the tangential stress of a pair is located between the atoms that carry it rather than at their positions. The estimator provides both $P_T^{\mathrm{IK}}(z)$ and $Q(z)$ from a PAV run, and the two-component films of Sec. IV.J show that the mixture form of the closure, with the conformally scaled $g(r)$ of the majority species for the cross and minority pairs, keeps the accuracy of the pure case, with a modest degradation for a near-monolayer of particles twice the size of the solvent. In those films the contour transfer at the midplane happened to be small, 0.15% of $\gamma$, because the two faces differ only by a thin adsorbed layer and $Q$ nearly cancels between them; it was nevertheless resolved and predicted to within 0.0004, and it would be larger for a thinner film, a film with a dense adsorbed layer or a liquid layer on one side, a supported film, or whenever the dividing plane cannot be placed at the centre. That the per-interface tensions obtained in this way return the tension of the pure film for a clean face and, at low coverage, an adsorbate surface pressure close to that of a two-dimensional ideal gas is, in our view, the strongest evidence that the local IK partition of a film without a bulk interior is physically meaningful. We regard this as the main application of the method.

*Per-atom binning and stability criteria.* Section II.E implies a practical rule for the mechanical-isotropy criterion of thin films [8,9]: the normal component must come from the IK or method-of-planes computes (or simply be set to the measured vapor pressure, which is what mechanical equilibrium demands), while the tangential component may come from the PAV profile provided it is corrected with (11). Both the uncorrected PAV $\Delta P_T$ and the IK $\Delta P_T$ identify the same films as confined and the same as bulk-like, because the correction vanishes wherever the interior is homogeneous, and both give the same surface tension (Table 1); but the magnitude of the anisotropy in the confined regime, and therefore any threshold placed on it, differs by 22–28% between the two. The IK value is the one that corresponds to a well-defined, locally equilibrated stress and should be the one reported.

# VI. CONCLUSIONS

The difference between the Irving–Kirkwood tangential pressure profile of a planar fluid and the profile obtained by binning per-atom virials (PAV), which is the Harasima profile, is the derivative of a pair quantity $Q(z)$ whose per-pair contribution is a linear ramp between the two atoms. Under the closure $\rho^{(2)} \simeq \rho(z_1)\rho(z_2)g(r_{12})$, with $g$ interpolated between its liquid and dilute limits according to the local density of each pair, $Q$ is a double integral of the measured density profile against a kernel built once from $u'(r)g(r)$. Adding $\mathrm{d}Q/\mathrm{d}z$ to the PAV tangential profile yields an estimate of the IK profile whose residual, for 20 Lennard-Jones films spanning thicknesses of 7–25$\sigma$, cutoffs of 3.5–7.5$\sigma$ and temperatures of 0.72–0.90, is 0.001–0.002 $\epsilon\sigma^{-3}$ overall, close to the statistical noise in the liquid interior, and captures 86–93% of the difference at the interface. The central-plane anisotropy that enters mechanical-stability criteria for thin films is reproduced to within 0.0012 over a range of 0.1, where the uncorrected PAV value is low by up to 28%; the surface tension is identical for the IK, PAV and estimated profiles. Applied to the normal component, the same closure removes 83–92% of the spurious structure of the per-atom $zz$ profile, providing a test of the method that needs no IK run. The local IK profile, and the term $Q(z)$ that the estimator provides with it to within 0.6% of $\gamma$, are what is needed when the question is not the total tension but its partition between two non-equivalent interfaces of a thin film, where the contour choice moves up to 10% of $\gamma$ from one interface to the other, or any other local stress criterion; the method makes these accessible from per-atom virial runs at no additional cost. On two-component films with a surface-active adsorbate unequally distributed between their faces, the mixture form of the closure keeps a comparable accuracy, the two interfacial tensions are resolved, the clean face of a singly decorated film recovers the tension of the pure film, the contour transfer between the faces is predicted to within 0.0004, and the surface pressure of the small adsorbate is that of a two-dimensional gas at low coverage. The method requires no additional pass over the film trajectory and removes the main computational cost of IK profiles in large-cutoff simulations.

# SUPPLEMENTARY MATERIAL

The supplementary material contains a description of the reference implementation `gauge.py` and its usage, the LAMMPS inputs for the bulk $g(r)$ runs and for the two-component films, the residual statistics by region, the flatness test and the accuracy of $Q$ for each of the 20 single-component films, the figures for the two-component films I and II, and the direct test of the instability of the homogeneous fluid at the mean density of the cells; the accompanying data archive contains `gauge.py`, the LAMMPS inputs and $g(r)$, the tabulated profiles and summaries of the 23 films, and the raw profiles of the three two-component films.

# ACKNOWLEDGMENTS

This study received financial support from an internal grant (18379) provided by the Universidad Michoacana de San Nicolás de Hidalgo.

# AUTHOR DECLARATIONS

**Conflict of interest.** The author has no conflicts to disclose.

# DATA AVAILABILITY

The data that support the findings of this study are available within the article and its supplementary material.